\documentclass[final, 5p, times]{elsarticle}
\usepackage{subcaption}
\usepackage{hyperref}
\usepackage{amsmath}
\usepackage{adjustbox}
\usepackage{xcolor}
\usepackage{makecell}
\journal{arxiv.org}

\begin{document}
\begin{frontmatter}

\title{Search for dark matter Particles via Invisible Decays in ${}^{46}$Sc Nuclear $\gamma$ Cascades with a CsI(Tl) Detector}

\author[inst1]{Sharada Sahoo}
\author[inst1]{Jing-han Chen}
\author[inst1]{Mahdi Mirzakhani}
\author[inst3]{Harikrishnan Ramani}
\author[inst1]{Rupak Mahapatra}
\author[inst2]{Surjeet Rajendran}

\affiliation[inst1]{organization={Department of Physics and Astronomy, Texas A\&M University },
            city={College Station},
            postcode={77840}, 
            state={TX},
            country={US}}

\affiliation[inst3]{organization={Physics and Astronomy, University of Delaware}, 
            city={Newark, Delaware},
            postcode={19716}, 
            country={US}}

\affiliation[inst2]{organization={Department of Physics and Astronomy, Johns Hopkins University}, 
            city={Baltimore, Maryland},
            postcode={21218}, 
            country={US}}

\begin{abstract}

 % In this study, we propose a novel high-statistics experiment to search for invisible decay modes in nuclear  $\gamma$ cascades using approximately 100 kg of CsI (Tl) scintillators available at Texas A\&M. The experiment aims to detect missing $\gamma$/energy by rigorously establishing the absence of a  $\gamma$ in a well-identified  $\gamma$ cascade, which could indicate the presence of any dark-sector particles such as Axions and Axion-Like Particles (ALPs), Dark Scalar, Dark Photon and Milli-Charged Particles. A high-intensity  $\gamma$ source, such as Sc46, positioned in the center of the detector, enables the search for ALPs with axion-photon coupling in the range of $10^{-8} < g_{a\gamma \gamma} < 10^{-2}$ and a mass range of $60~\mathrm{keV}$ to $1~\mathrm{MeV}$. This methodology indicates that the experiment has the potential to constrain previously unexplored parameter space not addressed by existing laboratory-based experiments.

Dark matter remains one of the most compelling open problems in modern physics, motivating experimental searches for new light, weakly coupled particles beyond the Standard Model. Despite extensive efforts employing diverse detection strategies, large regions of parameter space remain unexplored. We report a high-statistics laboratory search for invisible decay modes in nuclear  $\gamma$-ray cascades using approximately $100~\mathrm{kg}$ of CsI(Tl) scintillators operated at Texas A\&M University. The experiment employs a high-activity ${}^{46}$Sc radioactive source and a ``missing-$\gamma$'' technique, in which the absence of a photon from a well-identified cascade serves as a signature of new physics. Unlike appearance--disappearance experiments, this approach requires only a single photon conversion into a dark-sector particle, enabling sensitivity to significantly weaker couplings. The setup provides simultaneous sensitivity to a broad class of light dark-sector candidates, including axions and axion-like particles, dark scalars, and dark photons in the $0.1 - 1 \text{ MeV}$ mass region. Through careful control of detector containment, energy resolution, and environmental backgrounds, we exclude certain regions on the previously explored parameter space. With foreseeable improvements in detector volume and systematic uncertainty control, this technique has the potential to probe currently unexplored parameter space for axion-like particles and light dark scalars.

\end{abstract}

\end{frontmatter}

\section{Introduction}

Various astrophysical observations based on the behavior of galaxies in the galactic clusters \cite{Zwicky1,Zwicky2}, rotational velocity curves of stars in galaxies \cite{Rubin1}, gravitational lensing \cite{alcock}. have confirmed the existence of non-luminous matter in the universe or the so-called dark matter. About 83\% of the matter content of the universe is in the form of dark matter. This experiment aims to search for several dark matter candidates, including axions and axion-like particles (ALPs), dark scalars, dark photons, and milli-charged particles.

% Light, weakly coupled particles below the MeV scale arise naturally in many extensions of the Standard Model and are compelling candidates for the dark sector, either as dark matter itself or as mediators of its interactions. They also appear in frameworks addressing fundamental problems such as the strong CP problem, the gauge hierarchy, and the cosmological constant. Moreover, they have been invoked in astrophysical contexts, for example in energy transport mechanisms inside stars and as potential drivers of supernova shock revival. These motivations make the search for such particles a central goal of modern particle and astroparticle physics.  

% Astrophysical and cosmological observations have already placed strong bounds on light, weakly coupled particles. However, these limits are highly model-dependent and can often be avoided with modest modifications to the underlying theory. For instance, particles that couple predominantly to baryons remain largely unconstrained above $\sim 100$ keV, since they are too heavy to affect horizontal-branch stars yet may not efficiently cool supernovae if their interaction strength exceeds that of neutrinos ~\cite{Green_2017}. Similarly, cosmological bounds weaken significantly once the baryon abundance drops below the GeV scale or when density-dependent effects are introduced. Given these uncertainties, laboratory-based probes with well-controlled systematics are indispensable for establishing robust and complementary constraints.  

Light, weakly coupled particles below the MeV scale arise naturally in many extensions of the Standard Model and are compelling dark-sector candidates, either as dark matter itself or as mediators of its interactions. They also appear in solutions to the strong CP problem~\cite{PhysRevD.16.1791, Weinberg1978,Wilczek1978}, the gauge hierarchy problem~\cite{Graham_2015}, and the cosmological constant problem~\cite{Graham2019}, and have been invoked in astrophysical contexts such as stellar energy transport and supernova shock revival~\cite{Mori:2021pcv}. While astrophysical and cosmological observations place strong bounds on these particles, the limits are highly model-dependent and can be evaded with modest theoretical modifications. These considerations highlight the importance of laboratory-based searches with controlled systematics to provide robust and complementary probes of light, weakly coupled particles.

In this work, we present a high-statistic, laboratory-based search for rare invisible decay channels in well-identified nuclear $\gamma$ cascades. The experiment employs a ${}^{46}$Sc radioactive source placed at the center of a $\sim 100$ kg CsI(Tl) scintillator array. By exploiting coincidence tagging of the 1120~keV $\gamma$ line and searching for the associated 889~keV photon, the setup is designed to identify missing $\gamma$ events that may signal the conversion of a photon into a hidden-sector particle such as an axion-like particle, dark photon, or scalar mediator. Through dedicated simulations, careful control of detector containment, and extensive background characterization using passive shielding and active vetoes, this work demonstrates the feasibility of probing rare processes in a controlled environment and sets the stage for next-generation ton-scale experiments with improved containment and sensitivity.  

This paper is organized as follows. Section~2 presents the theoretical framework, encompassing the production of axions and axion-like particles (ALPs) from radioactive nuclear transitions, the resulting missing-$\gamma$ probability in the detector, and the overall sensitivity of the search. Section~3 describes the experimental setup, including the choice of radioactive source, detector assembly, passive and active shielding, and detector stability and characterization. Section~4 discusses the background sources relevant to the missing-$\gamma$ signal and quantifies their contributions. Section~5 presents the preliminary analysis performed with a $^{22}$Na source, followed by the missing-$\gamma$ signal extraction with the $^{46}$Sc source and the associated statistical and systematic uncertainties. Section~6 summarizes the exclusion limits obtained for various dark-sector candidates, including axions and axion-like particles, dark scalars, dark photons, and milli-charged particles, and outlines potential upgrades and future improvements to extend the sensitivity to previously unexplored regions of parameter space.

\section{Theoretical Framework} \label{Theory}

Nuclear $\gamma$ cascades provide a clean and well-controlled environment to search for exotic decay channels. Radioactive isotopes such as $^{60}\mathrm{Co}$ and $^{46}\mathrm{Sc}$ undergo beta decay followed by sequential $\gamma$ emissions as the daughter nucleus de-excites to its ground state. The resulting $\gamma$ spectra exhibit well defined and characteristic lines associated with each transition. Both isotopes feature two $\gamma$ lines, allowing one $\gamma$ to be used as a tag while the second is examined for anomalies. In practice, the large energy separation between the two $\gamma$ lines in $^{46}\mathrm{Sc}$ compared to $^{60}\mathrm{Co}$ makes $^{46}\mathrm{Sc}$ the preferred choice.

\begin{figure}[h!]
    \centering
    \includegraphics[width=0.7\linewidth]{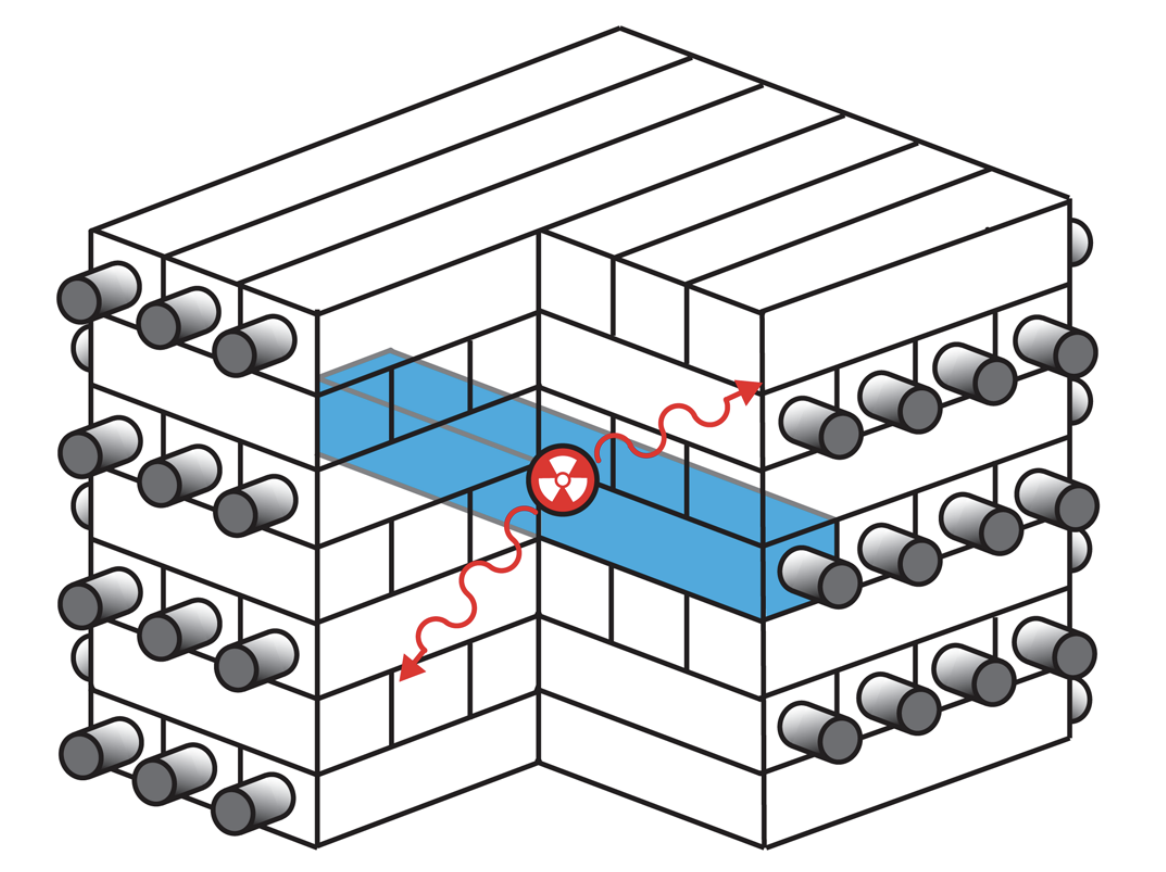}
    \caption{Schematic diagram of the proposed ton-scale stacked CsI detector array designed to efficiently contain  $\gamma$ rays emitted from a radioactive source located at the center of the detector \cite{Missing_Gamma2}. The emitted  $\gamma$ rays interact within the CsI crystal assembly, producing scintillation photons that are collected by photomultiplier tubes (PMTs) coupled to the crystals.}
    \label{fig:experiment_design}
\end{figure}

There are two generic mechanisms by which radioactive decays can produce dark sector particles. In the first, an intermediate excited nuclear state decays invisibly by emitting a dark sector particle instead of a $\gamma$ photon. In this case, the standard $\gamma$ cascade appears incomplete. In the second mechanism, a $\gamma$ emitted in the cascade subsequently converts into a hidden sector particle through suppressed but non zero interactions with surrounding nuclei or electromagnetic fields. Examples include axions or axion-like particles produced via Primakoff type processes or related conversion channels. In both scenarios, the final state consists of one observed $\gamma$ and one beyond Standard Model particle, leading to an apparent missing $\gamma$ in an otherwise well understood two $\gamma$ cascade. This missing $\gamma$ signature provides a direct and experimentally robust handle to search for new physics.

\begin{table}
\caption{Missing branching fractions of the dark matter candidates.}
\begin{adjustbox}{max width=0.5\textwidth}
    \begin{tabular}{|l|c|c|}
        \hline
        \textbf{Model} & $\mathcal{L}_{\text{int}}$ & $\text{BR}_{\text{miss}}$ \\
        \hline
        Scalar (nucleon coupling) 
        & $g_p \,\phi\, \bar{N} N$ 
        & $\dfrac{g_p^2}{2e^2} \left( 1 - \dfrac{m_\phi^2}{\omega^2} \right)^{5/2}$ \\
        \hline
        Dark Photon 
        & $\epsilon\, F^{\mu\nu} F'_{\mu\nu}$ 
        & $\epsilon^2 \left( 1 - \dfrac{m_{A'}^2}{\omega^2} \right)^{5/2}$ \\
        \hline
        Milli-charged Particle 
        & $-Q\, \bar{\chi}\gamma^\mu A_\mu \chi$ 
        & $Q^2 \,\dfrac{25\alpha}{9}\,\dfrac{\kappa(\omega,m_Q)}{\omega^5}$ \\
        \hline
        ALP (photon coupling) 
        & $\dfrac{1}{4 f_{a\gamma}}\, a\, F_{\mu\nu}\tilde{F}^{\mu\nu}$ 
        & $\dfrac{\sigma_P}{\sigma_P + \sigma_{SM}} \Big(1 - e^{-\ell/\lambda}\Big)$ \\
        \hline
    \end{tabular}
\end{adjustbox}
    \label{tab:branching_fractions}
\end{table}

This work focuses on a set of theoretically well-motivated models together with their effective interaction operators ~\cite{Missing_Gamma1}. Table~\ref{tab:branching_fractions} summarizes the expressions for the missing branching fractions corresponding to different dark matter candidates. The parameter $g_p$ denotes the effective scalar coupling to nucleons~\cite{Missing_Gamma2}, while $\phi$ represents a scalar mediator of mass $m_\phi$. The symbol $\epsilon$ refers to the kinetic mixing parameter between the Standard Model photon and the dark photon fields, where the dark photon has mass $m_{A'}$. The parameter $Q$ corresponds to the fractional electric charge of a milli-charged particle $\chi$ with mass $m_Q$. Finally, $a$ denotes an axion-like particle (ALP) whose coupling to photons is characterized by the decay constant $f_{a\gamma}$. The coupling constants $g_p$, $\epsilon$, $Q$, and $f_{a\gamma}$ thus define the primary experimental parameters governing the sensitivity of the search to new physics beyond the Standard Model.

This approach offers an advantage over other laboratory-based axion searches, such as those using reactors or beam-dump experiments or light shining through the wall experiments. In our case, we focus solely on the production of rare particles at the source, without relying on their subsequent conversion back to photons or electrons at the detection site. As a result, the detection strategy requires only a single interaction during the nuclear decay itself thereby increasing the probability of observing such rare events.

\section{Experimental set-up}
Completing the experiment requires both the construction and detailed characterization of the detector. The proposed setup consists of a $\sim$100 kg CsI(Tl) scintillator array with a radioactive source placed at its center. The source is selected such that it undergo decays that produce at least two particles: one acting as a trigger or tag, and the other providing sensitivity to potential rare or invisible decay modes.  

The choice of radioactive isotope is therefore crucial. Common candidates such as $^{24}\mathrm{Na}$, $^{60}\mathrm{Co}$, and $^{46}\mathrm{Sc}$ emit two primary  $\gamma$ rays suitable for coincidence measurements. Among these, $^{46}\mathrm{Sc}$ is particularly advantageous, as it emits well-separated  $\gamma$ lines at 889~keV and 1120~keV that can be clearly resolved by the detector. Its half-life of 84 days ensures a sufficiently long data-taking period, while the source activity can be tuned to provide high decay statistics without introducing excessive pile-up. These features make $^{46}\mathrm{Sc}$ a well-motivated choice for use with the CsI(Tl) detector. However, a key limitation of CsI(Tl) is its slow decay time ($3 \mu s$), which restricts the choice of source activity to $<37~\mathrm{kBq}$ ($<1~\mu$Ci) to prevent pile-ups in the data acquisition system.

\subsection{Detector Assembly}

CsI(Tl) crystals were selected for their high density (4.51~g/cm$^3$) and large scintillation light yield ($\sim 5.4 \times 10^4$ photons/MeV), with a peak emission wavelength of 550~nm. It also provide efficient containment of MeV-scale $\gamma$ rays and excellent energy resolution~\cite{VINCENT2023168624}. The detector array consists of CsI(Tl) crystals originally commissioned for the CLEO experiment~\cite{CLEO}, with more than 10~tons of material available for future detector expansions and ton-scale implementations of the missing-gamma technique.

\begin{figure}[h!]
    \centering
    \includegraphics[width=0.9\linewidth]{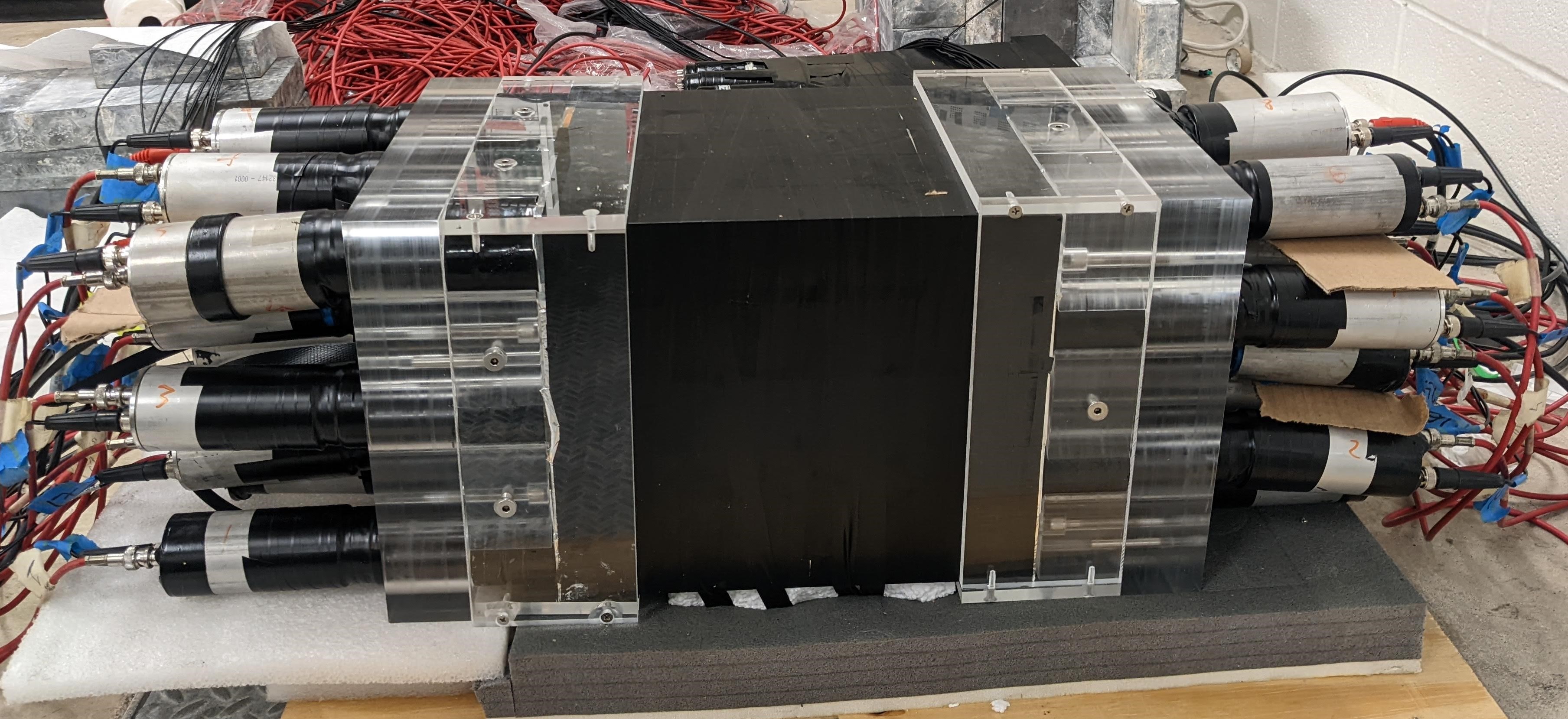}
    \caption{Photograph of the $\sim$100 kg CsI(Tl) detector assembly, serving as a first-generation proof-of-concept prototype for the proposed ton-scale detector design as shown in Fig.~\ref{fig:experiment_design}. The detector consists of a stacked crystal array coupled to photomultiplier tubes (PMTs), with alternating PMT orientations adopted to optimize packing geometry and reduce readout crowding. The central detector module is divided into fixed and movable sections, allowing radioactive sources to be inserted or replaced without disassembling the detector array.}    \label{fig:Detector_Assembly}
\end{figure}

Each CsI(Tl) crystal has dimensions of $30.5 \times 5.1 \times 5.1~\mathrm{cm^3}$ and a mass of approximately 4 kg. Crystals were selected after inspection for radiation damage, hygroscopic degradation, and mechanical defects~\cite{Verma}. Surface restoration and optical treatment were performed using progressively finer grit sandpapers followed by polishing liquids to improve surface smoothness, optical transparency, and scintillation light collection uniformity. Each crystal was subsequently wrapped with a high-reflectance PTFE layer providing $>92\%$ reflectivity for CsI(Tl) scintillation photons~\cite{PTFE_Tape}, followed by an outer aluminized Mylar layer for optical isolation and electromagnetic shielding~\cite{Al_Myler,VINCENT2023168624}. These preparation procedures ensure efficient collection and optical isolation of the scintillation photons produced within each crystal.

The wrapping fully covers the crystal surfaces except for one end, which is optically coupled to a photomultiplier tube (PMT) using EJ-550 silicone optical grease~\cite{OpticalGlue_response}. The optical coupling compound provides near-unity transmission efficiency at the CsI(Tl) scintillation emission wavelength, maximizing photon transport from the crystal into the PMT photocathode. Scintillation photons incident on the photocathode of the PMT are converted into photoelectrons and subsequently amplified through a multi-stage dynode chain present in the PMT, producing an electrical pulse proportional to the deposited energy. The detector employs 5~cm diameter Electron Tube 9954B series PMTs, whose photocathode area closely matches the crystal cross section. The PMTs provide a quantum efficiency of approximately 10\% at the 550~nm emission wavelength of CsI(Tl). Combined with the high scintillation light yield of the crystals, this results in sufficient photoelectron statistics to achieve excellent energy resolution, as demonstrated in Sec.~\ref{fig:CsI_Characterization}~\cite{PMT_response}. The rear end of each PMT is connected to a voltage-divider base, which distributes the supplied high voltage across the photocathode and dynode stages through SHV (Safe High Voltage) coaxial cables. The amplified PMT output signal is extracted from the base and transmitted to the digitizer using coaxial BNC cables for waveform acquisition and analysis.

The detector assembly consists of 26 CsI(Tl) crystals arranged in a $5 \times 5$ geometry. The central crystal is divided into fixed and removable sections to accommodate the radioactive source at the detector center, with the fixed section made slightly larger to support the surrounding crystals when the removable section is extracted. The crystal stack is held in a custom acrylic support structure that ensures uniform alignment and stable optical coupling. Adjacent crystals are oriented with alternating PMT directions to minimize crowding and facilitate readout integration. Cylindrical openings in the holder accommodate the PMT assemblies. The complete detector assembly is shown in Fig.~\ref{fig:Detector_Assembly}.

%Section about whole assembly, pmt, wires.%

%3.3 section should be added for detector characterisation.% 

%The complete setup consists of 26 CsI (Tl) crystals (including 2 central halves), arranged in a \(5 \times 5\) square matrix to form a compact and symmetric detection volume (Fig.\ref{fig:Detector_Assembly}). To suppress cosmic muon-induced background, a BC-400 plastic veto layer was added surrounding the scintillator array. Surrounding the entire detector and veto assembly, a 10~cm thick lead shield was installed to reduce the environmental  $\gamma$ background.  Additionally, the setup is enclosed in a thermal insulation box maintained at a constant temperature of $\approx 21\,^{\circ}$C to ensure stable PMT performance and minimize gain drift due to temperature fluctuations.% 

\subsection{DAQ system}
\label{subsec:daq_system}

% \begin{figure}
%     \centering
%     \includegraphics[width=0.8\linewidth]{ASCID_Figure/ASCID_setup.png}
%     \caption{Side view of the detector assembly: The design consists of 25 CsI crystals stacked in a $5 \times 5$ configuration. Surrounding the crystal array are plastic scintillator panels serving as a background veto, followed by a 10~cm lead shielding layer for additional background reduction. The entire setup is enclosed within a thermal enclosure, with external cooling provided to maintain optimal operating conditions for the PMTs.}
%     \label{fig:detector_design}
% \end{figure}

% \begin{figure}[h!]
%     \centering
%     \includegraphics[width=0.8\linewidth]{ASCID_Figure/Experimental_setip.png}
%     \caption{The full experimental setup includes the complete detector assembly, a high-voltage (HV) power supply delivering HV to the PMTs, and a digitizer that records scintillation signals from the PMTs for data acquisition and analysis.}
%     \label{fig:experiment_setup}
% \end{figure}

High-voltage biasing for the PMTs is provided by a CAEN SY5527LC mainframe equipped with two A7236 HV modules~\cite{HV_SY5527LC}. Each A7236 module provides 24 independent high-voltage channels with output capability up to 3.5 kV and 1.5 mA per channel~\cite{HV_A7236}, matching the operating requirements of the Electron Tubes 9954B PMTs, which typically operate between 1600--2200 V with current draw below 1 mA. Detector high voltage, channel currents, and system monitoring are controlled through the GECO2020 interface~\cite{GECO2020}.

Signal acquisition is performed using a CAEN V1740D waveform digitizer installed in a compact CAEN VME mini-crate~\cite{CAEN_V1740D}. The digitizer provides 64 input channels with LEMO connectors and operates at a sampling rate of 62.5 MHz, corresponding to a 16 ns time resolution. This sampling rate is sufficient to accurately resolve the $\sim3~\mu$s scintillation decay profile of CsI(Tl) crystals~\cite{GammaData2024_CsITlNa_DataSheet}. Data acquisition and online monitoring are handled using the CoMPASS software package~\cite{COMPASS}, which provides real-time waveform processing and energy spectrum visualization. 

Owing to the long CsI(Tl) scintillation decay time, a waveform acquisition window of $10~\mu$s was used, with the trigger position set at 40\% of the acquisition window. Event energies were reconstructed by integrating the pulse waveform over a $3~\mu$s interval beginning at the trigger time. CoMPASS performs online pulse integration and energy reconstruction during acquisition, eliminating the need to store individual waveforms and substantially reducing the data volume during detector commissioning and calibration runs. For offline analysis, reconstructed event information, including channel number, trigger timestamp, and deposited energy was recorded in Comma-Separated Values (CSV) format.

\subsection{Passive Shielding and Active Veto}
\label{sec:background_and_shielding}

The detector assembly incorporates dedicated passive shielding and an active veto system to suppress background radiation that could mimic a missing-$\gamma$ signature. The design and performance of these systems are described below. A comprehensive characterization of all the background sources relevant to the present search, including their estimated contributions to the missing-$\gamma$ signal region, is presented in Sec.~\ref{sec:Background_estimate}.

% The dominant background arises from the finite  $\gamma$-ray containment of the detector. The ${}^{46}\mathrm{Sc}$ source is placed at the center of a $\sim100\,\mathrm{kg}$ CsI(Tl) detector, providing a minimum path length of approximately 10~cm for each emitted  $\gamma$. However, the absorption efficiency of CsI(Tl) at 889\,keV and 1120\,keV is less than unity. Consequently, a fraction of otherwise standard cascade events will lose one  $\gamma$ due to escape from the active volume, producing an irreducible background. The detector containment efficiency is evaluated using detailed \textsc{GEANT4} simulations by computing the fraction of events depositing energy within the region of interest (ROI) (Described briefly in the Sec.~\ref{sec:monte_carlo}).

Background radiation include intrinsic radioactive contaminants in the CsI(Tl) crystals and external sources such as ambient  $\gamma$ rays, $\beta$ particles, cosmic-ray--induced particles, and neutrons. $\gamma$ rays with energies near the tagging line ($\sim1120~\mathrm{keV}$) are particularly relevant, as they can add false count to the missing gamma signal. To suppress external backgrounds, the detector assembly is enclosed within a lead shield constructed from $20.4 \times 10.2 \times 5.1~\mathrm{cm^3}$ lead bricks, providing a minimum shielding thickness of $10.2~\mathrm{cm}$ on all six sides. Lead shielding of this thickness attenuates $\sim1$ MeV  $\gamma$ rays by approximately three orders of magnitude~\cite{Lead_Shielding}.

In addition, 11 plastic scintillator panels, each measuring $35.6 \times 10.2 \times 5.1~\mathrm{cm^3}$, are deployed as an active veto layer surrounding the detector inside the lead shielding. The scintillators are fabricated from BC-408 plastic scintillator material previously used in the CDMS-II experiment~\cite{Plastic_Scintillator,PhysRevLett.106.131302}. The veto system is designed to reject residual backgrounds not sufficiently attenuated by the passive lead shielding, particularly cosmic-ray--induced events. Each scintillator panel was machined, polished, and wrapped following procedures similar to those used for the CsI(Tl) crystals. PMTs coupled to the scintillators provide veto signals through the readout channels. Events were selected by requiring activity in the CsI(Tl) detector array with no coincident signal in the veto scintillators within the predefined coincidence window~\cite{sahoo2026reactorbasedsearchaxionlikeparticles}. 

A dedicated background measurement was performed using this configuration to characterize the residual environmental and cosmic-ray backgrounds. The corresponding analysis is presented in Sec.~\ref{sec:data_analysis}, and the resulting background energy spectrum is shown in Fig.~\ref{fig:Background}.

\subsection{Stability and Thermalization}

Prior to extended data collection with the ${}^{46}\mathrm{Sc}$ source, we conducted detector stability studies by monitoring the temporal variation of the calibration peak positions. During the initial stages of operation, we identified a gradual downward drift of the $1275~\mathrm{keV}$ calibration peak, which we attributed to time-dependent gain variations in the detector response. Further investigation revealed that sustained operation induced heating of the PMT bases, whose temperature-dependent behavior modifies the signal current and consequently alters the ADC response of the detector channels~\cite{LI2024, PMT_response}. To characterize and mitigate this effect, we implemented an active cooling scheme while continuously recording the PMT-base temperature via thermocouples. A thermal enclosure was constructed around the detector assembly to suppress heat exchange with the ambient environment, through which temperature-regulated air was circulated by means of an external cooling system. The temperature of the injected air was stabilized near $21^\circ\mathrm{C}$.

\begin{figure}[h!]
    \centering
    \includegraphics[width=0.8\columnwidth]{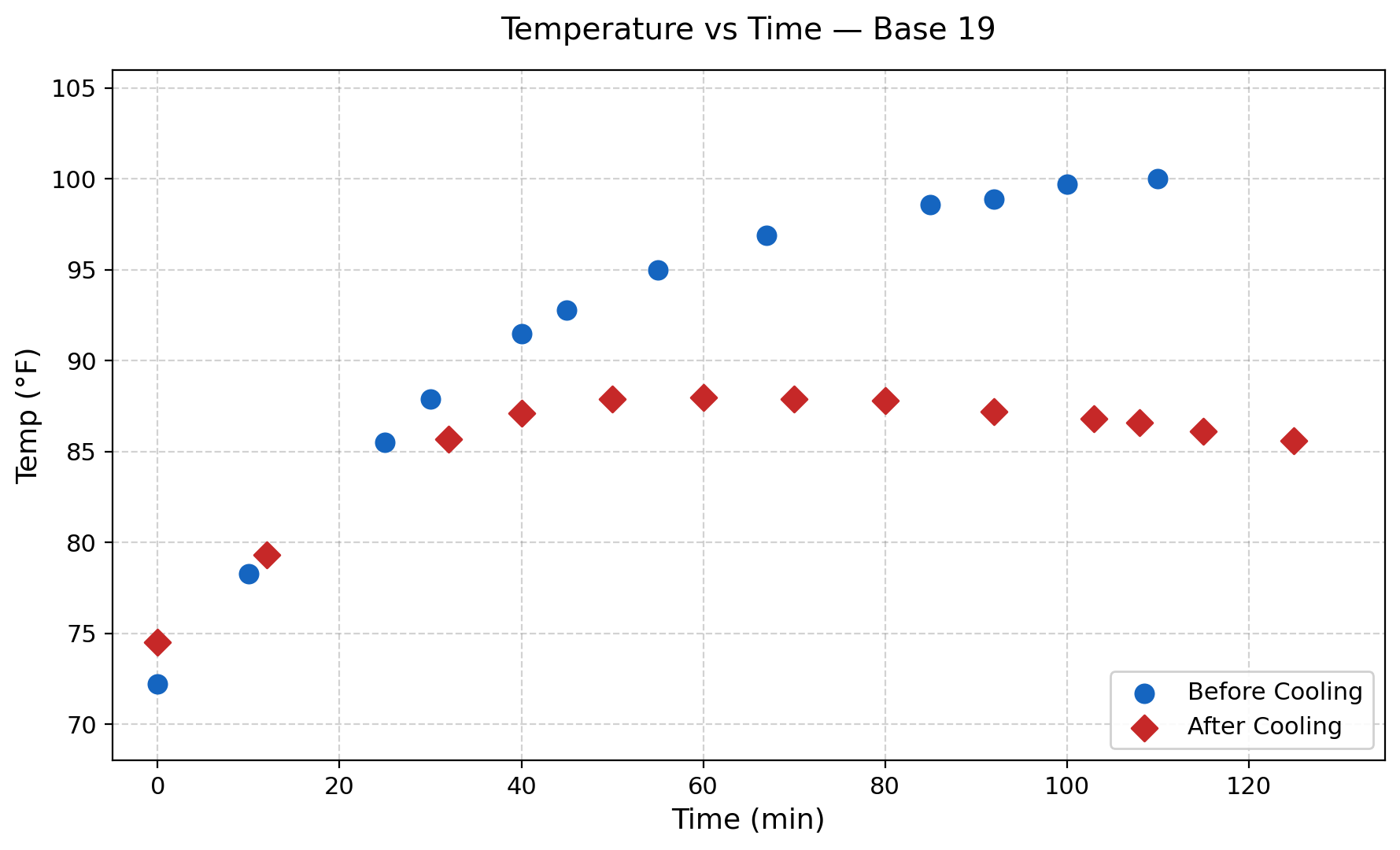}
    \caption{
    PMT-base temperature measured during detector operation before and after the application of external cooling. Without cooling, the temperature rises above \(38^\circ\mathrm{C}\), while active cooling stabilizes the operating temperature near \(31^\circ\mathrm{C}\), significantly improving detector stability during long-duration operation.
    }
    \label{fig:Cooling}
\end{figure}

\begin{figure}[h!]
    \centering
    \includegraphics[width=0.8\columnwidth]{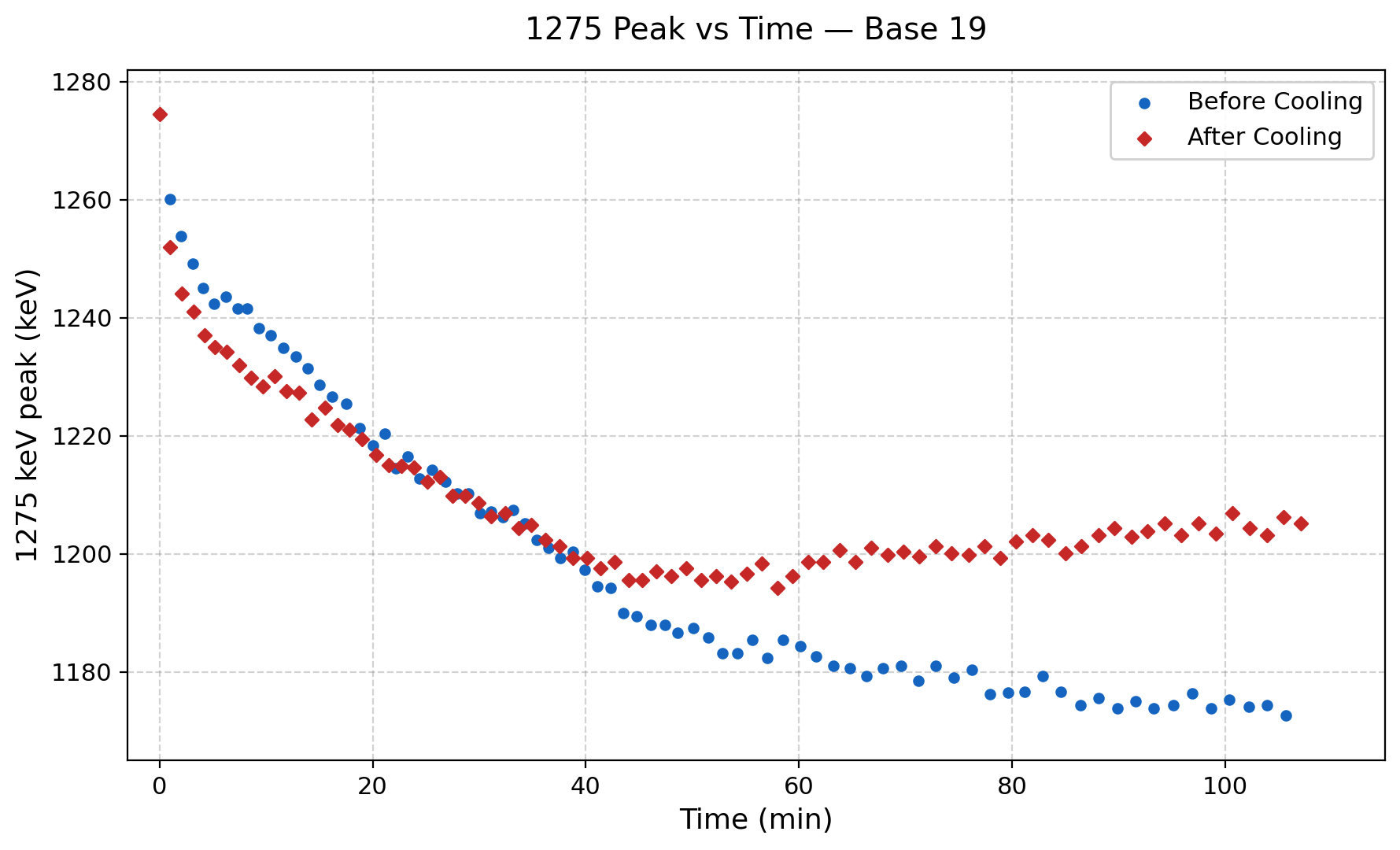}
    \caption{
    Temporal evolution of the \(1275~\mathrm{keV}\) calibration peak position for base-19. Blue points correspond to measurements taken before cooling, showing a continuous downward drift from approximately \(1274.5~\mathrm{keV}\) to \(1173~\mathrm{keV}\) over \(\sim106\) minutes. Red points correspond to measurements taken after cooling, showing significantly improved stability following an initial transient response.
    }
    \label{fig:Cooling_1275}
\end{figure}

With external cooling applied, the PMT-base temperature stabilized near \(31^\circ\mathrm{C}\), compared to temperatures exceeding \(38^\circ\mathrm{C}\) in the absence of cooling, as shown in Fig.~\ref{fig:Cooling}. The improved thermal stability substantially reduced calibration drift during extended operation, as illustrated in Fig.~\ref{fig:Cooling_1275}. Consequently looking into both the figures, all physics data runs were initiated only after a \(40\)-minute stabilization period following high-voltage turn-on to ensure stable detector response. The combination of active cooling and controlled operating conditions effectively mitigated PMT-base overheating and enabled stable long-term detector performance.

In addition, dedicated analysis procedures were implemented to preserve detector stability throughout the full data-taking period. To further compensate for the time-dependent ADC gain variations, each detector channel was independently re-calibrated in successive \(15\)-minute intervals. The choice of this calibration window was determined quantitatively by studying the temporal stability and uniformity of the calibration peak positions shown in Figs.~\ref{fig:Cooling} and~\ref{fig:Cooling_1275}. This procedure preserves the intrinsic detector resolution and maintains stable energy reconstruction during long-duration detector operation.

\subsection{Detector Characterization}

Several detector-characterization studies were performed to validate the performance of the experimental setup. These studies include measurements of energy linearity, energy resolution, and the effective trigger threshold of the data acquisition system.

\begin{figure}[h!]
\centering

\begin{subfigure}{0.49\columnwidth}
    \centering
    \includegraphics[width=\linewidth]{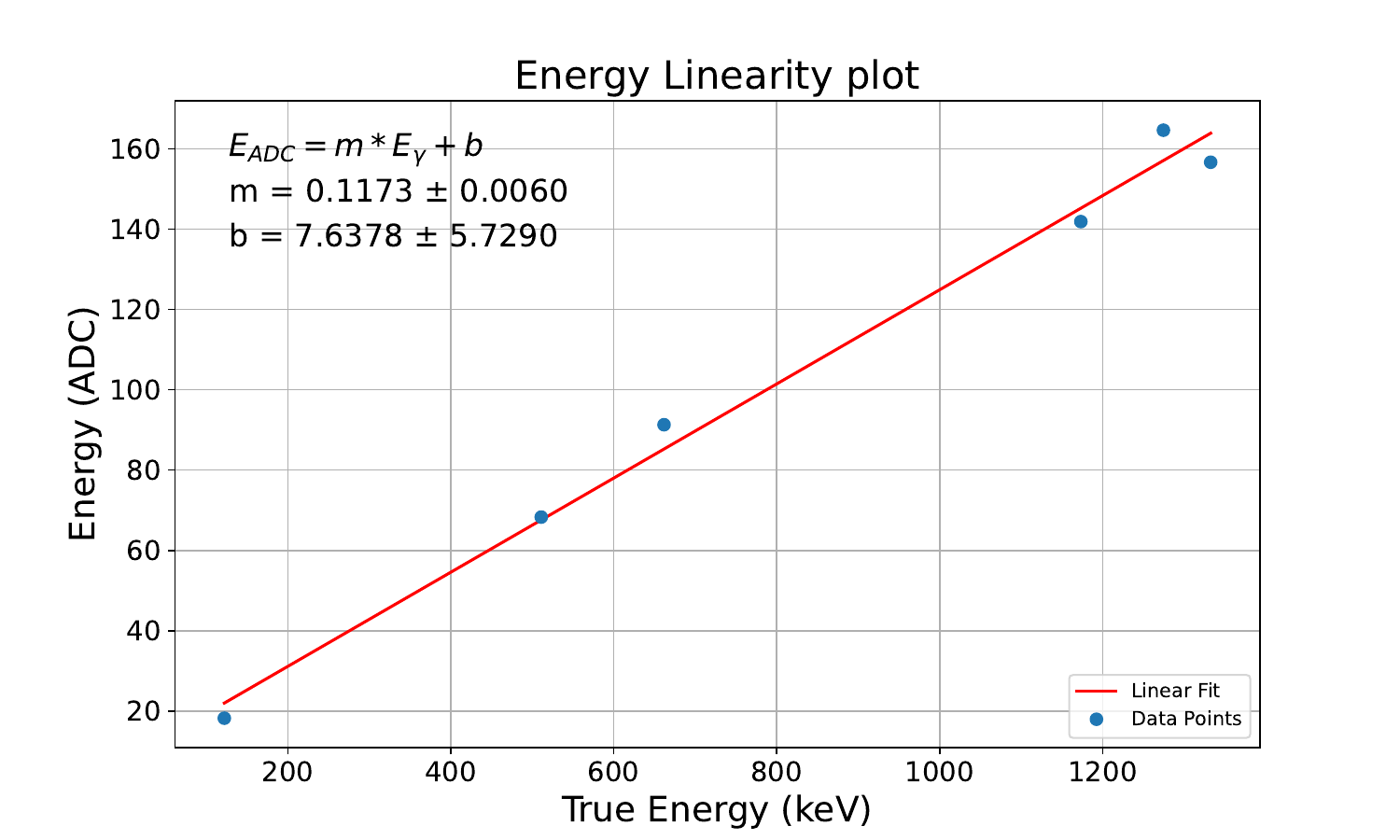}
    \caption{Energy linearity.}
    \label{fig:CsI_Tl_Energy_Linearity}
\end{subfigure}
\hfill
\begin{subfigure}{0.49\columnwidth}
    \centering
    \includegraphics[width=\linewidth]{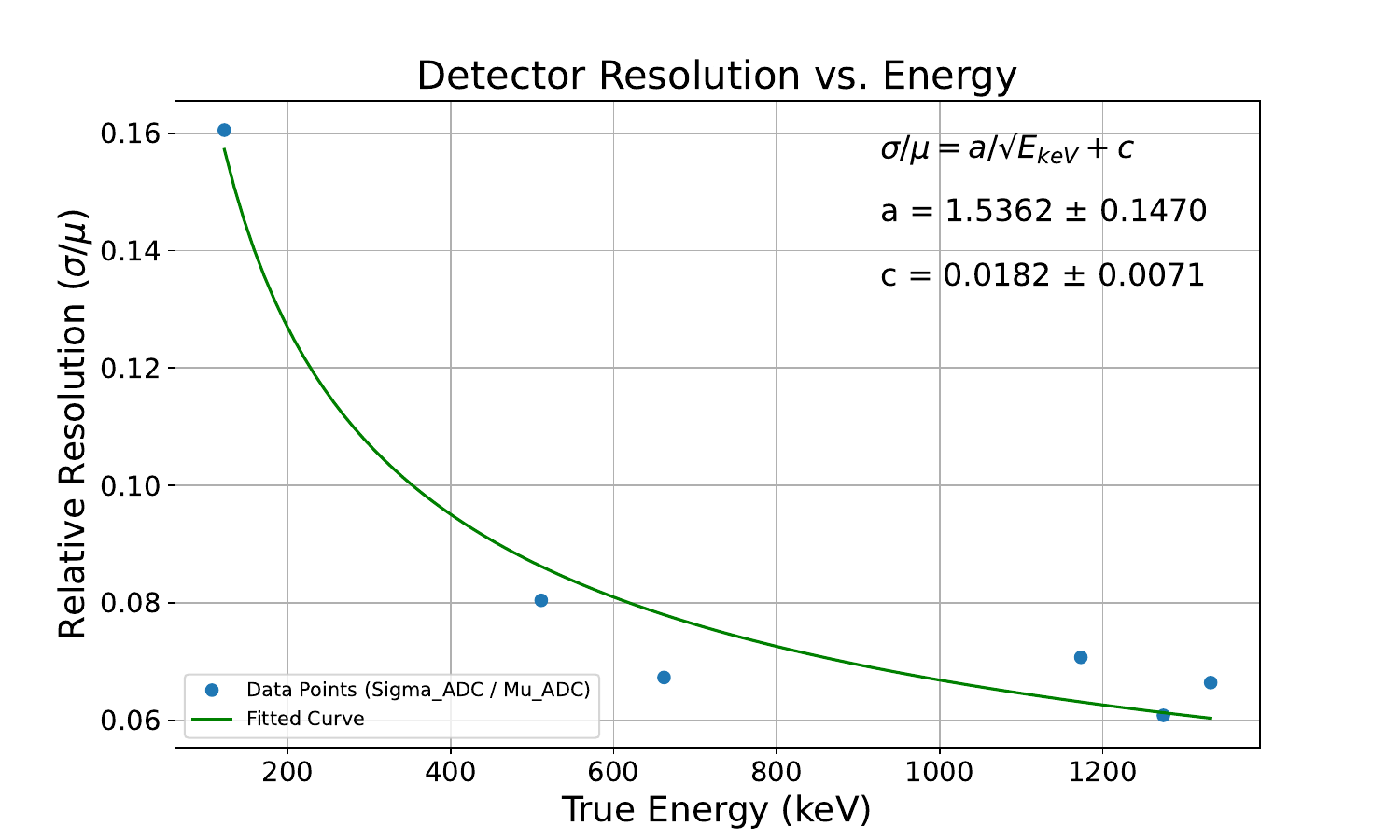}
    \caption{Energy resolution.}
    \label{fig:Energy_Resolution}
\end{subfigure}

\caption{
Characterization of a representative CsI(Tl) detector. 
Left: energy linearity measured using multiple radioactive calibration sources over the energy range \(100\text{--}1600~\mathrm{keV}\). 
Right: relative energy resolution, \(\sigma/\mu\), as a function of deposited energy, showing the expected scintillator scaling behavior.}
\label{fig:CsI_Characterization}
\end{figure}

Each CsI(Tl) crystal is characterized using radioactive sources such as ${}^{22}$Na, ${}^{60}$Co, ${}^{57}$Co, ${}^{137}$Cs, and ${}^{133}$Ba. $\gamma$-ray lines spanning the energy range $122$--$1460~\mathrm{keV}$ were measured with the sources placed directly along the long axis of the CsI(Tl) crystal at its midpoint, with scintillation photons collected at one end of the long axis by a coupled PMT. The measured ADC response as a function of deposited energy is well described by a linear calibration of the form~\cite{Energy_Linearity}
\begin{equation}
    E_\mathrm{ADC} = m \cdot E_\gamma + b,
    \label{eq:calib}
\end{equation}
where $m = (0.117 \pm 0.006)~\mathrm{ADC/keV}$ is the gain and $b = (7.64 \pm 5.7)~\mathrm{ADC}$ is the offset, the fit yields an excellent value of $R^2$ of 0.9897. The detector demonstrates 
stable and reliable performance of the crystal--PMT--base assembly for precision  $\gamma$-ray spectroscopy, as shown in Fig.~\ref{fig:CsI_Tl_Energy_Linearity}. 

The detector's energy resolution was also checked; it achieves an $\sigma/ \mu$ of approximately \(7\%\) around $1~\mathrm{MeV}$ (Fig.~\ref{fig:Energy_Resolution}), which is among the best values obtained for PMT-based CsI(Tl) systems. The measured energy resolution follows the expected scintillation-detector behavior,
\[
\frac{\sigma}{\mu} = \frac{a}{\sqrt{E}} + c,
\]
where \(E\) is the energy at which $\sigma$ and $\mu$ were calculated~\cite{Aprile_2020}. Here $a = (1.5362 \pm 0.147)$ and $b = (0.018 \pm 0.007)$, the fit yields an excellent value of $R^2$ of 0.9647. The observed behavior is well described by this relation over the full measured energy range, as illustrated in Fig.~\ref{fig:Energy_Resolution}.

The threshold efficiency of each detector channel was also evaluated to ensure reliable signal acquisition. For every channel, the digitizer threshold was set manually within the CoMPASS software~\cite{COMPASS} interfaced with the CAEN V1740D digitizer~\cite{CAEN_V1740D}. Hardware threshold values are expressed in units of least significant bits (LSB), which translate to an equivalent voltage in millivolts determined by the dynamic range and the selected energy resolution of the respective channel. The threshold for each channel was optimized to maximize the collection efficiency of signal $\gamma$-rays while simultaneously suppressing the dark-current pulses inherent to PMT operation; only pulses whose amplitude exceeds this threshold are accepted by the digitizer for further analysis. 

\begin{figure}[h]
    \centering
    \includegraphics[width=0.9\linewidth]{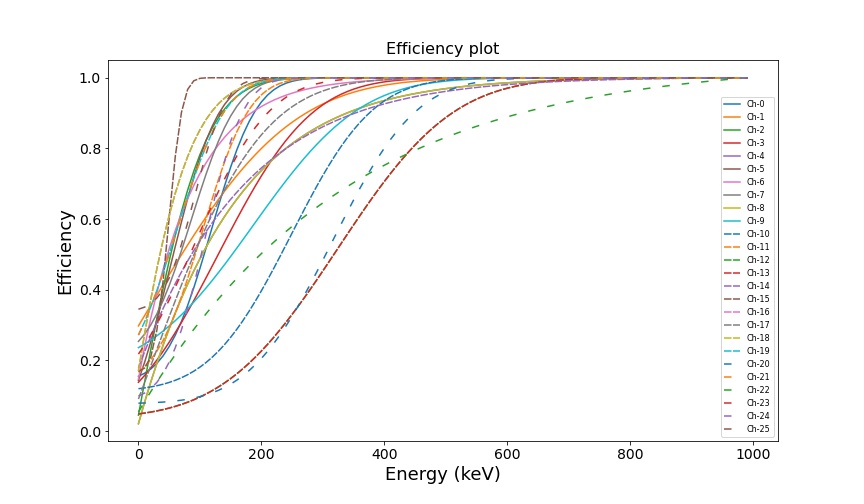}
\caption{Effective hardware trigger thresholds for all detector channels, exhibiting the expected error-function-like turn-on behavior~\cite{Threshold_efficiency}. The majority of channels reach full collection efficiency near $200~\mathrm{keV}$, well below the $889~\mathrm{keV}$ and $1120~\mathrm{keV}$ signal energies of the ${}^{46}\mathrm{Sc}$ cascade, confirming that the detector operates with high efficiency across the entire energy region of interest.}
    \label{fig:Energy_Threshold}
\end{figure}

To quantify the threshold efficiency, two energy spectra were acquired independently for each channel: one recorded at a threshold of $0$~LSB, which imposes no amplitude discrimination, and a second recorded at the operationally selected threshold value. The threshold efficiency as a function of energy was then obtained by computing the bin-by-bin ratio of the threshold spectrum to the zero-threshold reference spectrum. The resulting efficiency curves for all channels are presented in Fig.~\ref{fig:Energy_Threshold}.

The effective hardware trigger thresholds exhibit the expected error-function-like turn-on behavior~\cite{Threshold_efficiency}, with the majority of channels reaching full efficiency in the vicinity of $200~\mathrm{keV}$. A small subset of channels display elevated thresholds near $500~\mathrm{keV}$, while a few channels attain full efficiency at lower energies near $100~\mathrm{keV}$, as illustrated in Fig.~\ref{fig:Energy_Threshold}. To ensure complete collection of signal $\gamma$-rays across all channels, a common software threshold of $100~\mathrm{keV}$ was applied, thereby retaining all events above the lowest hardware threshold observed in the array.

Since the characteristic $\gamma$-ray energies from the \({}^{46}\mathrm{Sc}\) decay (\(889~\mathrm{keV}\) and \(1120~\mathrm{keV}\)) lie well above the threshold region, the detector operates with high efficiency for the signals of interest. These characterization studies provide confidence in the suitability of the detector system for the radioactive-source-based missing-\(\gamma\) search and associated dark-sector investigations.

\section{Data Analysis}
\label{sec:data_analysis}

A total of more than 300 hours of data were collected using a ${}^{46}$Sc radioactive source with the CsI(Tl) detector array and analyzed through the data analysis pipeline. 

The analysis pipeline processes the CSV output produced by the DAQ system (Sec.~\ref{subsec:daq_system}) and reconstructs individual decay events by clustering all triggers occurring within a $0.5~\mu$s coincidence window relative to the first triggered channel, provided that the recorded energy exceeds the software threshold of 100~keV (the determination of the coincidence window is discussed in Sec.~\ref{sec:coincidence_window}). A single decay event may therefore contain energy deposits recorded in multiple detector crystals.

From this event-building procedure, two key observables are constructed. The first is the event multiplicity, defined as the number of detector channels registering a trigger within a given decay event. Physically, this quantity reflects the number of correlated energy depositions arising from particles emitted in a single radioactive decay. The second is the co-added energy spectrum, defined as the sum of the energies deposited in all triggered crystals associated with the same decay event. These observables provide complementary information on the event topology and energy deposition pattern and form the basis of the missing-$\gamma$ search described in the following sections.

\subsection{Source-free Background Data Analysis}

\begin{figure}[h!]
\centering
\includegraphics[height=4cm]{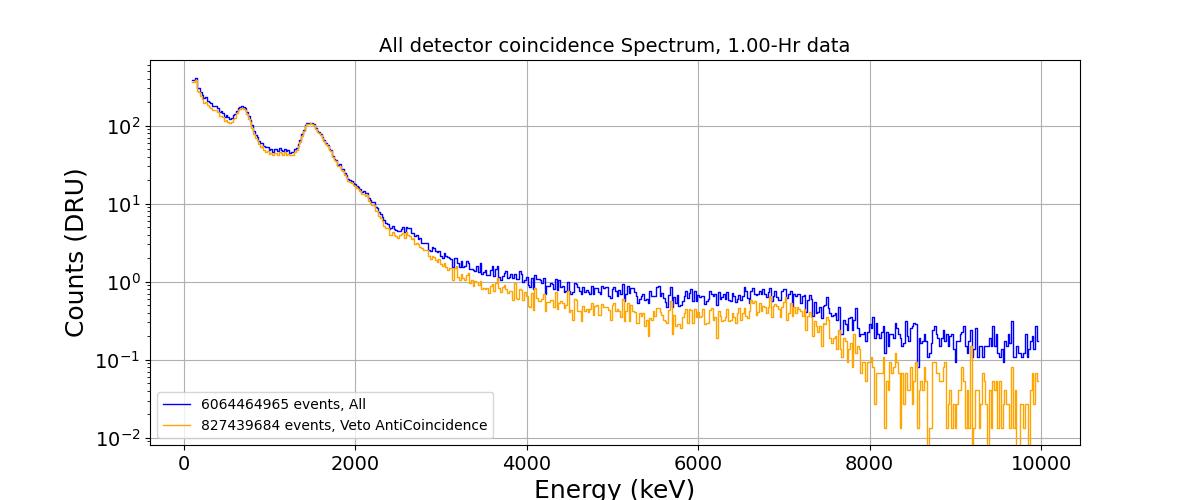}
\caption[Energy spectra from CsI(Tl) without any external source.]{Energy spectra recorded with the CsI(Tl) detector array in the absence of an external radioactive source, showing contributions from environmental radioactivity and intrinsic crystal backgrounds. The spectrum is shown over the energy range 0--10~MeV, where peaks associated with \({}^{137}\mathrm{Cs}\) (660~keV) and \({}^{40}\mathrm{K}\) (1460~keV) are visible. The blue curve corresponds to the full background spectrum, while the orange curve shows the spectrum after application of the active veto.}
\label{fig:Background}
\end{figure}

Data were first collected without any radioactive source to characterize the environmental background. The resulting energy spectrum, shown in Fig.~\ref{fig:Background}, serves as a reference for estimating the environmental background contribution to the rare-event search. Application of the active veto reduces the background event rate by more than 15\%, demonstrating the effectiveness of the veto system in suppressing cosmic-ray-induced and other residual background events.

\begin{figure}[h!]
\centering
\includegraphics[width=0.5\textwidth]{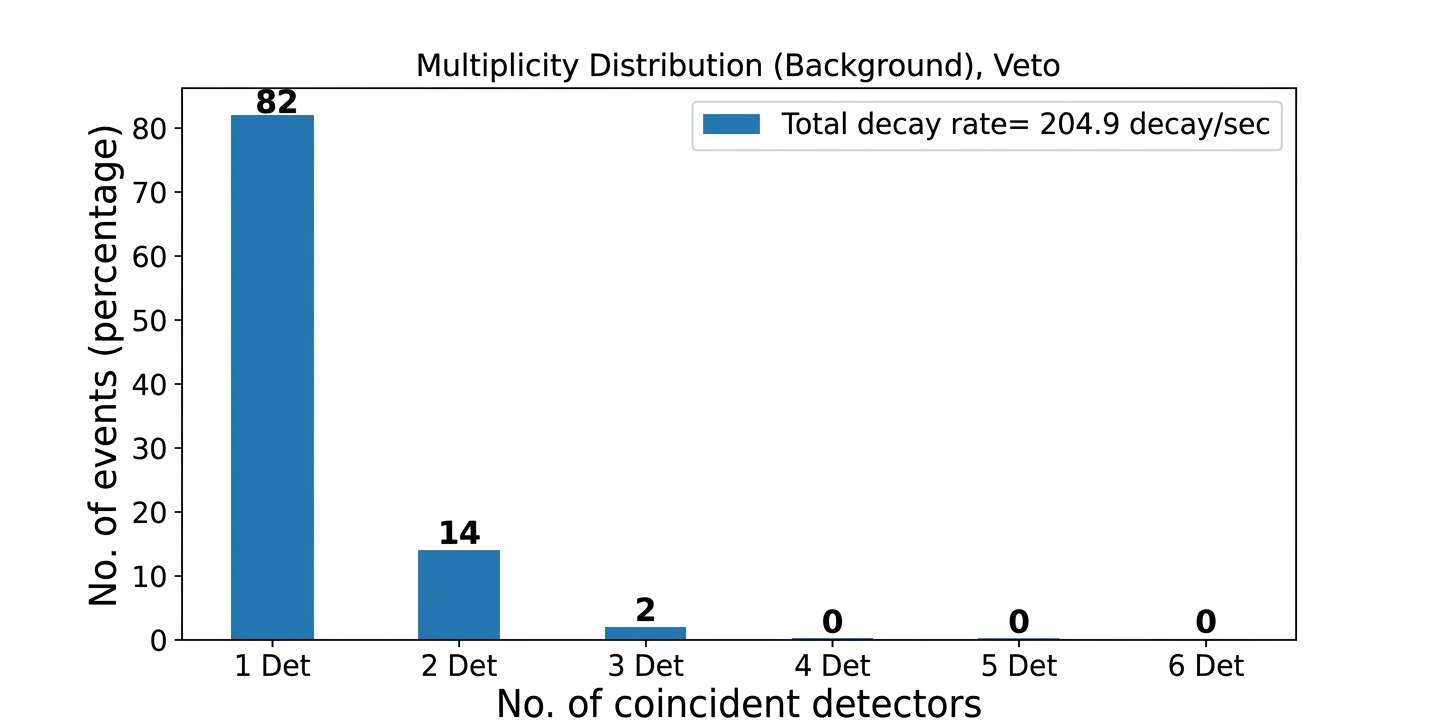}
\caption{Event multiplicity distribution recorded without an external radioactive source. The spectrum is dominated by multiplicity-one events, consistent with uncorrelated environmental background interactions.}
\label{fig:Multiplicity_Background}
\end{figure}

The event multiplicity distribution for the background dataset is shown in Fig.~\ref{fig:Multiplicity_Background}. As expected, the distribution is dominated by multiplicity-one events, reflecting the largely uncorrelated nature of environmental background interactions. Events with higher multiplicities arise primarily from Compton scattering of higher-energy $\gamma$ rays across multiple detector crystals. This distribution provides an important reference for comparison with source-induced events, which frequently produce correlated energy depositions in multiple detectors.

\subsection{Preliminary data analysis with ${}^{22}\mathrm{Na}$}
\label{sec:Na22:analysis}

The ${}^{22}\mathrm{Na}$ source was used to verify detector functionality, containment efficiency and analysis pipeline. In each ${}^{22}\mathrm{Na}$ decay, a positron is emitted in coincidence with a 1275~keV $\gamma$ ray. The positron subsequently annihilates with an electron, producing two additional 511~keV $\gamma$ rays, such that a single decay can yield up to three $\gamma$s. Although all three $\gamma$ rays are emitted nearly simultaneously on the nanosecond timescale~\cite{Cherry2012_NuclearMedicine}, the corresponding detector triggers are spread over longer times due to the relatively long decay time of the CsI(Tl) scintillator~\cite{GammaData2024_CsITlNa_DataSheet} and the trigger efficiency of the digitizer. This makes the choice of an appropriate time coincidence window a critical aspect of the analysis (discussed in Sec.~\ref{sec:coincidence_window}).

\begin{figure}[h!]
  \centering
  \includegraphics[width=0.5\textwidth]{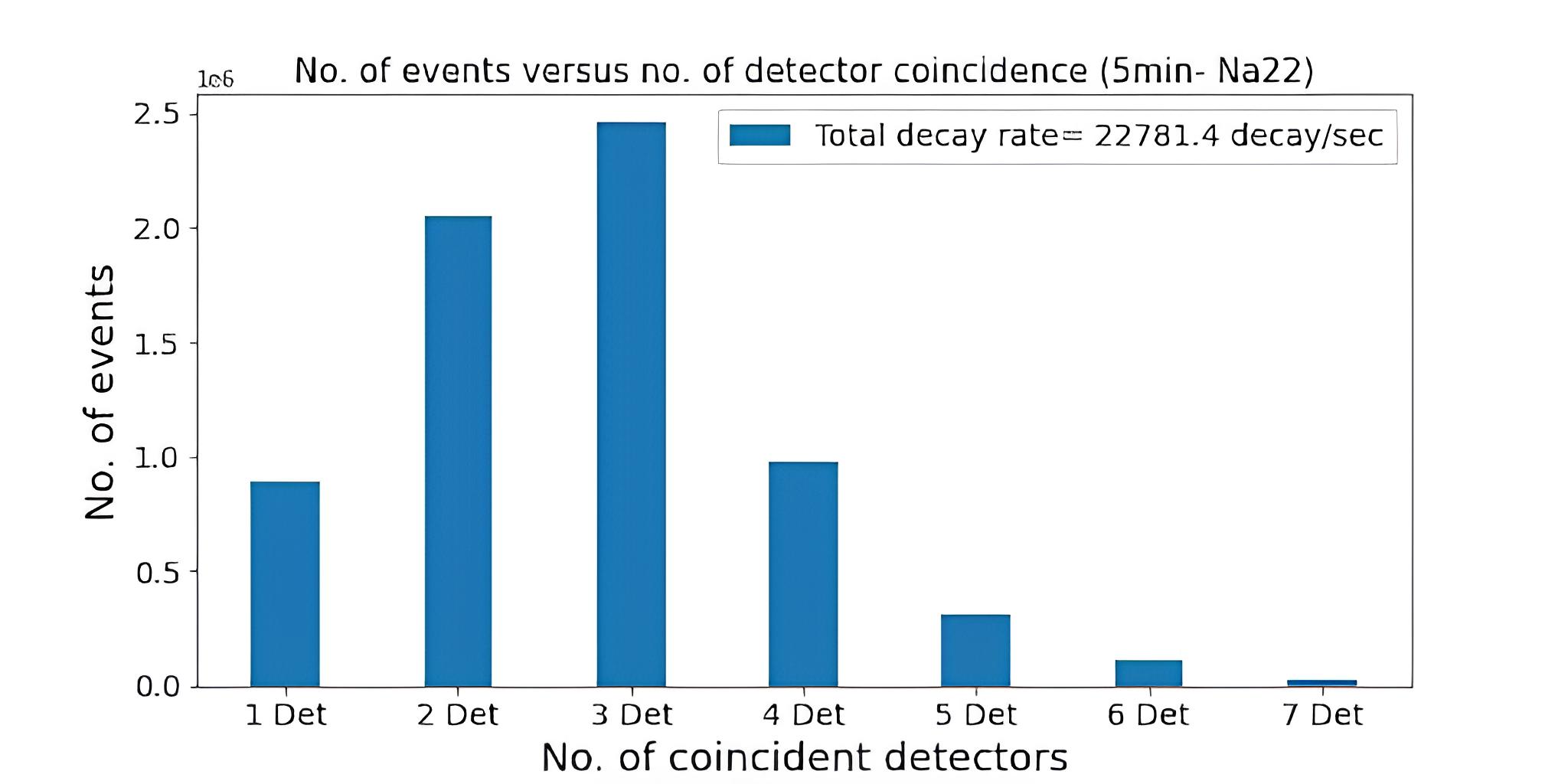}
    \caption{Event multiplicity distribution for the detector array with a ${}^{22}\mathrm{Na}$ source. The peak at multiplicity 3 corresponds to the three-particle decay signature of ${}^{22}\mathrm{Na}$, consisting of one 1275~keV $\gamma$ ray and two 511~keV annihilation $\gamma$ rays. Events with higher multiplicities arise primarily from Compton scattering of $\gamma$ rays across multiple detector crystals, while lower multiplicities result from incomplete energy containment, detector inefficiencies, accidental background events, or rare decay topologies.}
  \label{fig:Multiplicity_3_3}
\end{figure}

Event multiplicity is shown in the Fig.~\ref{fig:Multiplicity_3_3}. This distribution peaks at a multiplicity of three, consistent with the emission of three $\gamma$ rays in the ${}^{22}\mathrm{Na}$ decay cascade. In some cases, more than three signals are also observed when one or more $\gamma$ rays undergo Compton scattering within the detector. Conversely, fewer signals may be recorded if $\gamma$ rays escape without interaction or deposit energy below the trigger threshold. Events with fewer than the expected number of signals constitute potential signal candidates and are discussed in detail in later sections.

\begin{figure}[h!]
\centering
\begin{subfigure}{0.48\textwidth}
  \includegraphics[width=\linewidth]{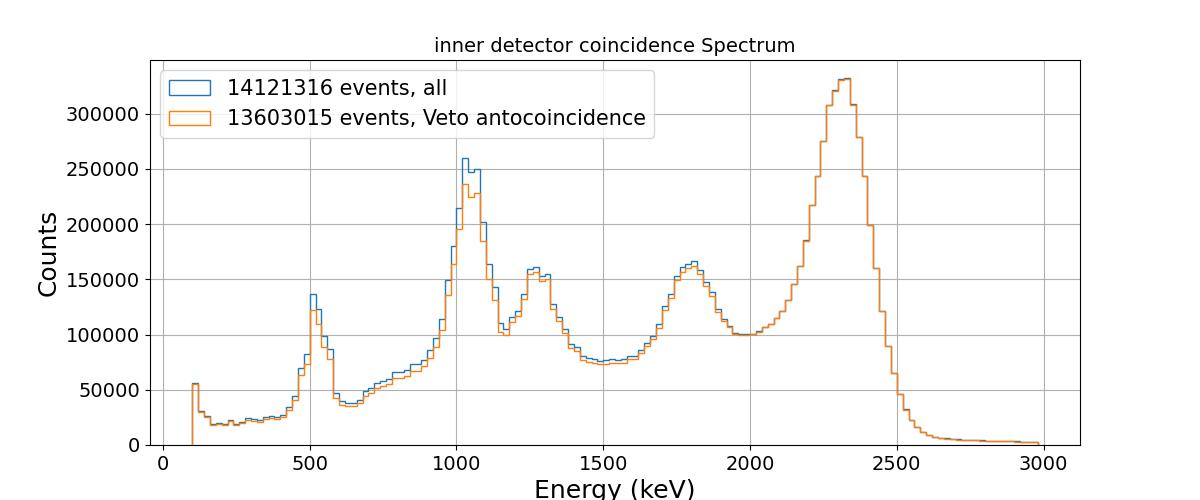}
  % \caption{.}
\end{subfigure}
\hfill
\begin{subfigure}{0.48\textwidth}
  \includegraphics[width=\linewidth]{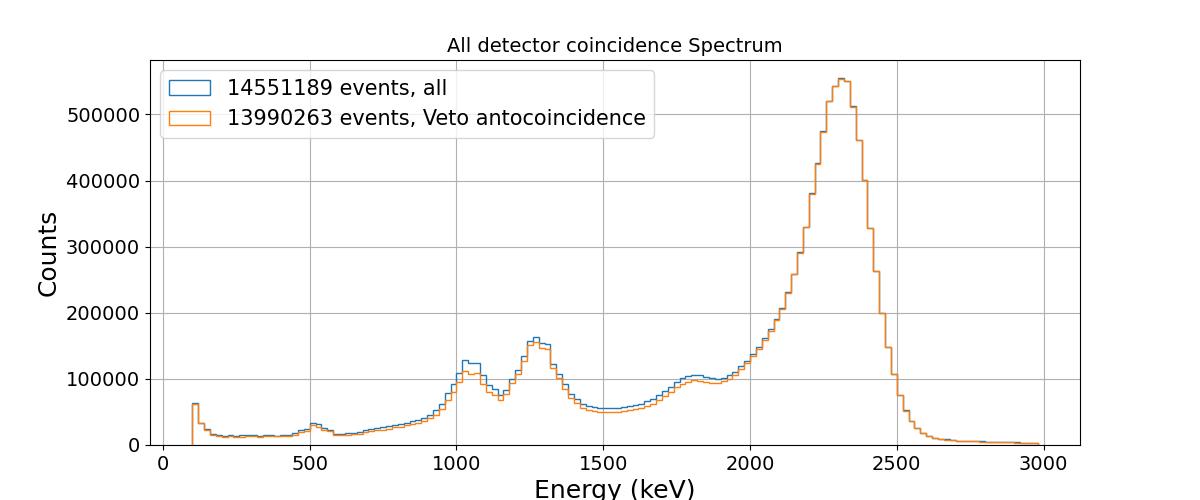}
  % \caption{.}
\end{subfigure}
\caption{Co-added energy spectrum of ${}^{22}\mathrm{Na}$ for the inner $3 \times 3$ detector channels (top) and the full $5 \times 5$ detector array (bottom). The orange curve represents the experimental data with the veto rejection cut applied, while the blue curve shows the spectrum without the veto cut. The $5 \times 5$ configuration exhibits a more pronounced full-energy absorption peak at 2297 keV, whereas the $3 \times 3$ configuration shows additional structures associated with partial energy containment and escape events.}
\label{fig:Co_added_Na22}
\end{figure}

In addition, the co-added energy spectrum as shown in the Fig.~\ref{fig:Co_added_Na22} shows good agreement with the expected spectrum, providing further confidence to proceed with measurements using the ${}^{46}\mathrm{Sc}$ radioactive source. The peak at $2297~\mathrm{keV}$ corresponds to full energy absorption of all three $\gamma$ rays ($1275 + 511 + 511~\mathrm{keV}$), while the remaining peaks arise from partial containment. The $511~\mathrm{keV}$ peak represents events in which only one annihilation $\gamma$ is detected; the $1022~\mathrm{keV}$ peak corresponds to detection of both $511~\mathrm{keV}$ $\gamma$s alone; the $1275~\mathrm{keV}$ peak arises either from escape of both annihilation $\gamma$s or from the alternate ${}^{22}\mathrm{Na}$ branch emitting a single $1275~\mathrm{keV}$ $\gamma$; and the $1786~\mathrm{keV}$ peak corresponds to escape of one $511~\mathrm{keV}$ $\gamma$. As seen in Fig.~\ref{fig:Co_added_Na22}, as the effective detector volume increases from the inner $3\times3$ to the full $5\times5$ array, escape peaks are suppressed and the total absorption peak becomes more prominent, consistent with improved $\gamma$-ray containment. Additional details of the ${}^{22}\mathrm{Na}$ studies are provided in the Appendix~\ref{app:Na22_analysis}.

% However, a significant fraction of the data was rendered unusable due to failures in the photomultiplier tube (PMT) bases. These failures mostly rise due to overheating of resistors in the base, which prevented proper biasing of the PMT and resulted in a complete loss of signal in the affected channel, and leading to unlikely energy value. To overcome this, the affected data was discarded and the faulty bases were replaced. After applying strict data quality selection criteria, only slightly more than $100$ hours of high-quality data were retained for the final analysis.

% Monte Carlo simulations of the experiment were performed using the \textsc{GEANT4} simulation toolkit developed by CERN, with assistance from a collaborator at Texas A\&M University. The simulations accurately reproduce the experimental geometry and incorporate the radioactive source, detector energy thresholds, and energy-dependent energy resolution effects. The \textsc{GEANT4} PhysicsList package is used to model particle interactions with matter, including electromagnetic interactions of electrons and photons with the matter. Additional details of the simulation framework are provided in the Appendix~\ref{sec:Appendices}.
\subsection{DAQ Validation for ${}^{46}$Sc }

\begin{figure}[h]
    \centering
    \includegraphics[width=8cm]{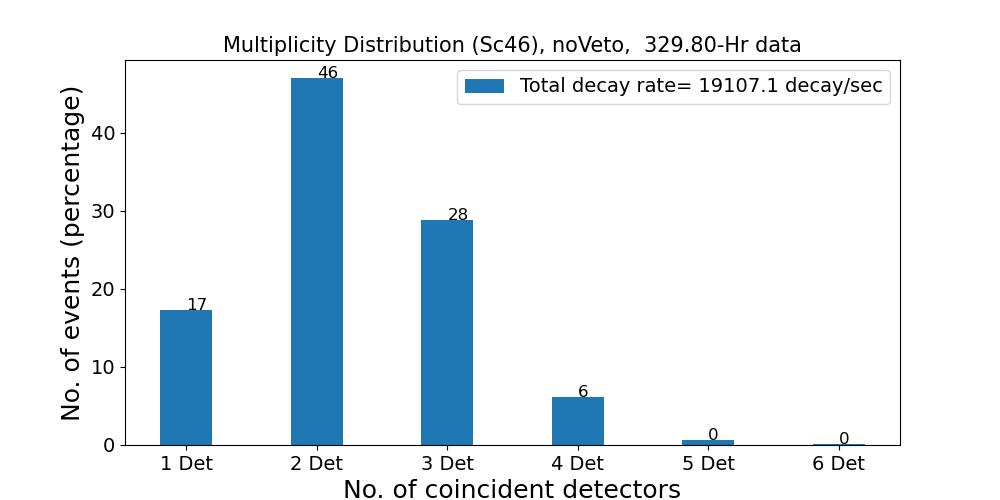}
    \caption{Event multiplicity distribution for the detector array with ${}^{46}$Sc source. The peak at multiplicity 2 corresponds to the two-particle decay of the radioactive source, while higher multiplicities arise from Compton scattering across multiple crystals.}
    \label{fig:Sc_Multiplicity}
\end{figure}

Using the defined coincidence window of \(0.5~\mu\mathrm{s}\) and the previously defined energy threshold of \(100~\mathrm{keV}\) for all detector channels, the event multiplicity distribution was constructed and is shown in Fig.~\ref{fig:Sc_Multiplicity}. The distribution peaks at a multiplicity of two, consistent with the expected emission of two  $\gamma$ rays in the ${}^{46}$Sc decay cascade. This agreement confirms again the proper functioning of both the detector system and the analysis framework. Events with multiplicity greater than two arise primarily from Compton scattering, in which a  $\gamma$ ray deposits energy in multiple detector channels. Events with multiplicity equal to one define the signal region and include contributions from detector containment losses, environmental backgrounds, and potential exotic decay signatures. The composition of this region is quantified in the following sections.

\subsection{Calculation of Missing $\gamma$ signal with ${}^{46}$Sc}

\begin{figure}[h]
    \centering
    \includegraphics[width=8cm]{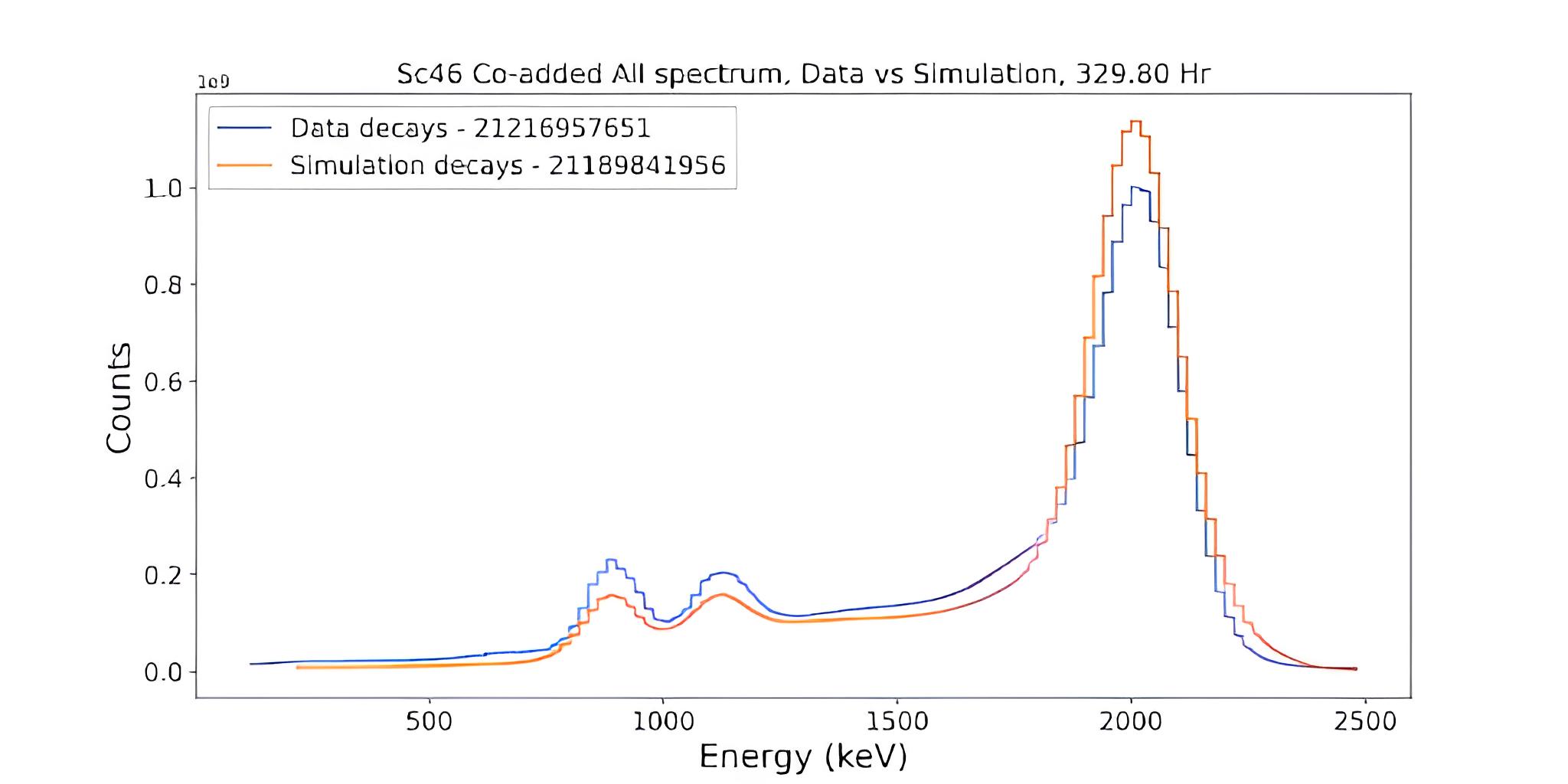}
    \caption{Co-added energy spectrum of ${}^{46}$Sc for all detector channels. The orange curve represents experimental data, while the blue curve shows the simulated spectrum. The peak near $\sim$2 MeV corresponds to the total decay energy, whereas lower-energy peaks indicate incomplete containment of the emitted radiation within the detector array. Note: The spectrum shown here is presented without background subtraction, where the measured ambient background spectrum is shown separately in Fig.~\ref{fig:Background}. }
    \label{fig:Sc_spectrumAll}
\end{figure}

To identify potential missing-$\gamma$ events, we examine the co-added energy spectrum, defined as the total energy deposited across all detector crystals within a single decay. The spectrum, shown in Fig.~\ref{fig:Sc_spectrumAll}, exhibits a clear total-absorption peak at approximately $2~\mathrm{MeV}$ ($889~\mathrm{keV} + 1120~\mathrm{keV}$) and single-$\gamma$ absorption peaks at $889~\mathrm{keV}$ and $1120~\mathrm{keV}$, corresponding to events in which one $\gamma$ ray is detected while the other escapes the detector volume. Good agreement is observed between data and Monte Carlo simulation, with the detector modelling and simulation validation described in Sec.~\ref{sec:monte_carlo}. The single-$\gamma$ peaks provide the baseline for the missing-$\gamma$ search, as discussed in the following sections.

For the present analysis, the $1120~\mathrm{keV}$  $\gamma$ ray is used as the tagging photon, and the search is performed for the coincident $889~\mathrm{keV}$  $\gamma$ ray. This choice is motivated by two considerations. First, the $889~\mathrm{keV}$ peak is susceptible to contamination from Compton-scattered $1120~\mathrm{keV}$ events; using the $889~\mathrm{keV}$ line as the tagging photon could therefore lead to misidentification of Compton-scattered $1120~\mathrm{keV}$  $\gamma$ rays as true cascade events. Second, the detector exhibits higher containment efficiency at $889~\mathrm{keV}$ than at $1120~\mathrm{keV}$. Taken together, these considerations make the $1120~\mathrm{keV}$ line the more reliable tagging channel for this analysis.

While Fig.~\ref{fig:Sc_spectrumAll} shows good overall agreement between data and simulation, discrepancies are observed in several regions, most notably at the $1120~\mathrm{keV}$ single-$\gamma$ absorption peak, which corresponds to events with a missing $889~\mathrm{keV}$ $\gamma$. The number of missing-$\gamma$ events is estimated by counting events within the $1120~\mathrm{keV} \pm 1.5\sigma$ region of the co-added energy spectrum across all channels, where the choice of $1.5\sigma$ is motivated by the separation between the $889~\mathrm{keV}$ and $1120~\mathrm{keV}$ peaks. For a total exposure of $7.54 \times 10^{9}$ ${}^{46}$Sc decays, the data yield $690.19 \pm 25.6 \times 10^{6}$ ($9.15\%$) missing-$\gamma$ events, while the simulation predicts $565.27 \pm 9.08 \times 10^{6}$ ($7.49\%$) events from standard processes, leaving an unaccounted excess of $125.08 \pm 27.2 \times 10^{6}$ ($1.65\%$) events beyond detector containment alone. The background contributions to this excess are evaluated quantitatively in Sec.~\ref{sec:Background_estimate}.

\subsection{Statistical and Systematic Uncertainty}
\label{sec:statistical_systematic}

The statistical uncertainty on the missing-$\gamma$ count is given by $\sigma_\mathrm{stat} = \sqrt{N_B}$~\cite{statistical_uncertainty}, where $N_B$ is the number of events in the signal region. For the present dataset, this yields a statistical uncertainty of $0.00035\%$, which is negligible owing to the large event statistics accumulated over the full data-taking period. Systematic uncertainties were evaluated using a cut-variation approach~\cite{systematic_uncertainty}, in which each analysis variable is varied around its nominal value and the resulting spread in the missing-$\gamma$ count is taken as the systematic uncertainty for that variable. The individual contributions are tested to be mutually independent and are therefore summed in quadrature to obtain the total systematic uncertainty. 

The total systematic uncertainty receives four significant contributions. (1)~\textbf{Coincidence window:} varying the window from $0.40$ to $0.60~\mu$s yields an uncertainty of $0.058\%$, (2)~\textbf{Energy threshold:} varying the software threshold from $60$ to $140~\mathrm{keV}$ gives an uncertainty of $0.267\%$. (3)~\textbf{Calibration shifts:} automated peak-finding can misidentify the true photo-peak position by $1$--$2$ ADC bins per channel; shifting each channel independently and repeating the full analysis yields an uncertainty of $0.21\%$. (4)~\textbf{Detector resolution:} the energy-resolution function $\sigma/\mu$ was sampled from Gaussian distributions defined by the fit covariance matrix and propagated through the simulation, yielding an uncertainty of $0.12\%$. Summing these contributions in quadrature gives a total systematic uncertainty of $0.373\%$. Table~\ref{tab:MissingGammaSummary} summarizes the missing-$\gamma$ fraction, the relevant background contributions, and the associated uncertainties. Further details of the uncertainty evaluation are provided in Appendix~\ref{sec:uncertainty}.

\begin{table*}[t]
\caption{Summary of missing $\gamma$ contributions and associated uncertainties. Numbers are given relative to the total $7.54 \times 10^9$ decays, with percentages in parentheses.}
\centering
\begin{tabular}{|l|c|c|c|}
\hline
\textbf{Category} & \textbf{Number ($\times 10^6$) (\%)} & \textbf{Systematic (\%)} & \textbf{Statistical (\%)} \\
\hline
Total Decay & 7542.8 (100) & -- & -- \\
\hline
Missing $\gamma$ events -- data (Tag 1120 keV) & 690.19 (9.15) & 0.34 & 0.00035 \\
\hline
Containment Inefficiency (Simulation) & 565.27 (7.49) & 0.12 & 0.00031 \\
\hline
Pile-up (in the ROI at 889 keV) & 70.53 (0.94) & 0.01 & 0.0001 \\
\hline
Radioactive background & 4.16 (0.055) & 0.001 & 0.00003 \\
\hline
BRICC (Conversion electron from a decay) & 1.21 (0.016) & theoretical & 0.00001 \\
\hline
\textbf{Remaining excess} & \textbf{49.04 (0.65)} & \textbf{0.373} & \textbf{$<$ 0.001} \\
\hline
\end{tabular}
\label{tab:MissingGammaSummary}
\end{table*}

\section{Background Estimation}
\label{sec:Background_estimate}

Despite the inherent advantages of the missing-$\gamma$ search strategy, several background processes can produce signatures indistinguishable from a genuine dark-sector decay. These arise from four primary sources: environmental and intrinsic radioactive backgrounds, which are partially suppressed by the passive shielding and active veto system described in Sec.~\ref{sec:background_and_shielding}; finite detector containment, which causes a fraction of signal $\gamma$-rays to escape the active volume undetected; DAQ-induced pile-up, in which a secondary pulse falling within the post-trigger window is misidentified as a missing-$\gamma$ event; and bound-state internal conversion via the BrICC process, which mimics an invisible nuclear transition. Each of these contributions and their combined effect on the missing-$\gamma$ signal region is summarized and quantified in the following Sec.~\ref{sec:Background_estimate}.

\subsection{Detector Containment}
Accurate evaluation of the detector containment efficiency is critical for estimating the fraction of missing $\gamma$-ray events prior to performing the experiment. As a first-order approximation, the missing fraction was estimated numerically using the detector geometry, the attenuation properties of CsI(Tl), and isotropically emitted \(889~\mathrm{keV}\) $\gamma$ rays propagating through the crystal array.

\subsubsection{Analytical Calculation of Containment}

\begin{figure}[h!]
    \centering
    \includegraphics[width=\columnwidth]{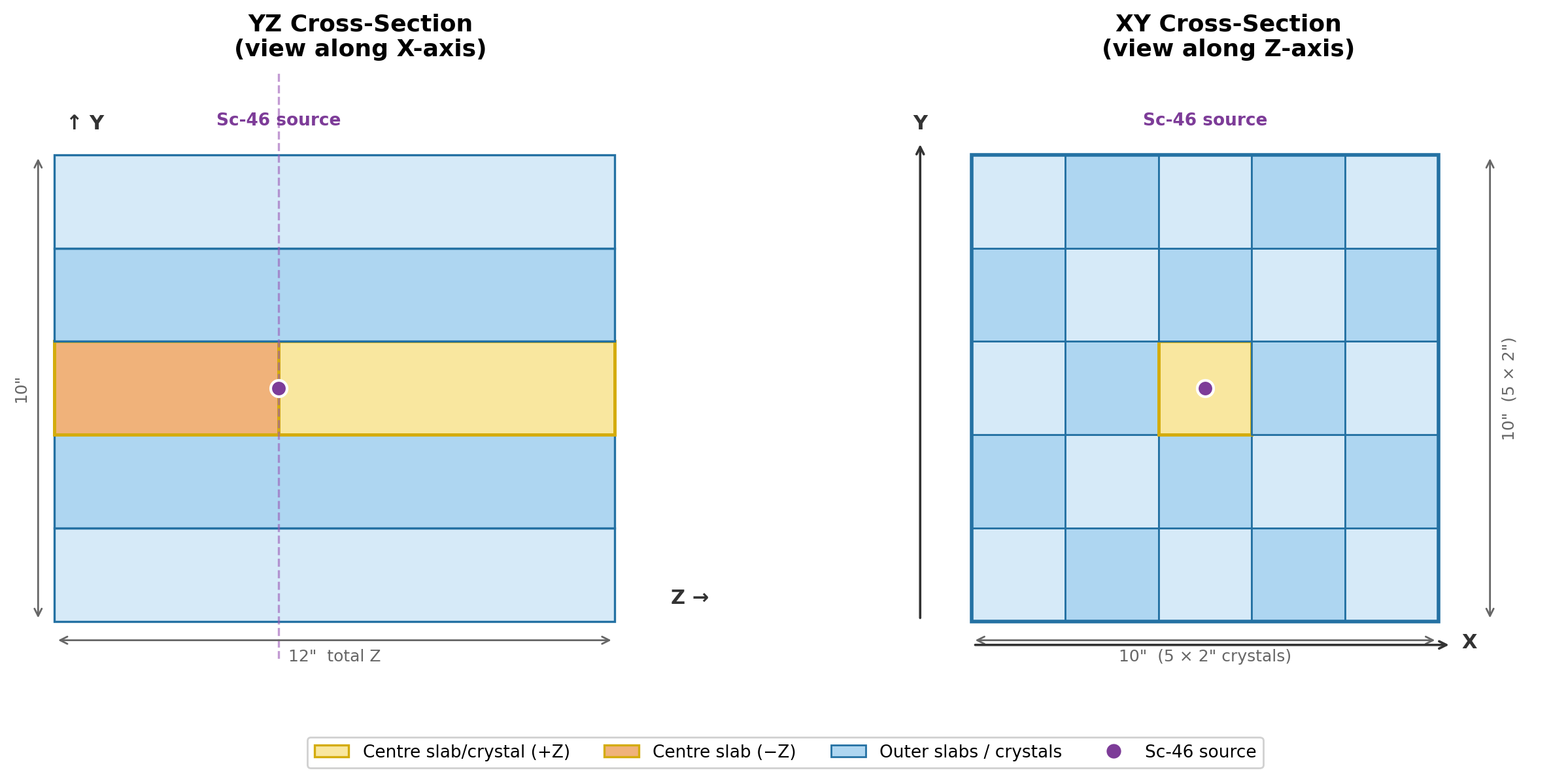}
    \caption{
    Cross-sectional views of the CsI(Tl) detector geometry with the \({}^{46}\mathrm{Sc}\) source located at the center. 
    Left: YZ cross-section (view along the X-axis) showing the stacked crystal arrangement along the vertical direction. 
    Right: XY cross-section (view along the Z-axis) showing the \(5\times5\) array of \(2''\times2''\) CsI(Tl) crystals surrounding the source position. 
    The central region is segmented symmetrically about the source to optimize containment and event reconstruction.
    }
    \label{fig:csi_geometry}
\end{figure}

Missing gamma signal is calculated with \(1132~\mathrm{keV}\) $\gamma$ ray being captured and \(889~\mathrm{keV}\) $\gamma$ ray being missing. The probability that an \(889~\mathrm{keV}\) $\gamma$ ray from \({}^{46}\mathrm{Sc}\) escapes the CsI(Tl) crystal array without depositing detectable energy was estimated using the Beer--Lambert attenuation law~\cite{Baur_2019},
\[
I = I_0 e^{-\mu x},
\]
where the total mass attenuation coefficient at \(889~\mathrm{keV}\) is \(\mu/\rho = 6.132\times10^{-2}~\mathrm{cm}^2/\mathrm{g}\), taken from the NIST XCOM database~\cite{NIST_XCOM}. Using the CsI(Tl) density \(\rho = 4.51~\mathrm{g/cm^3}\), this corresponds to a linear attenuation coefficient of \(\mu = 0.2766~\mathrm{cm}^{-1}\), or an attenuation length of \(\lambda = 3.62~\mathrm{cm}\). A Monte Carlo integration over isotropically distributed \(4\pi\) emission directions through the rectangular crystal geometry shown in Fig.~\ref{fig:csi_geometry} yields a geometric missing fraction of (the calculation is shown in further detail in Sec.~\ref{app:missing_containment_analytical})
\[
f_{\rm miss} = 2.773\%.
\]

This estimate accounts only for events in which the $889~\mathrm{keV}$ $\gamma$ escapes the detector without interacting. Applying the same calculation to the $1120~\mathrm{keV}$ $\gamma$ yields an absorption probability of $95.7\%$. The combined probability of fully absorbing the $1120~\mathrm{keV}$ $\gamma$ while the $889~\mathrm{keV}$ $\gamma$ escapes is therefore $0.957 \times 0.0277 = 2.65\%$. This is an order-of-magnitude estimate of the missing-$\gamma$ fraction due to incomplete containment; the actual value may differ substantially because of effects not captured here. In particular, partial energy deposition via Compton scattering of one or both cascade $\gamma$s, together with the finite detector resolution and energy threshold, can shift this result. A quantitative treatment would require a full detector-response simulation as explained in the following section.

\subsubsection{Monte-Carlo Estimation of the Containment}
\label{sec:monte_carlo}

\begin{figure}[h]
  \centering
  \includegraphics[width=0.9\columnwidth]{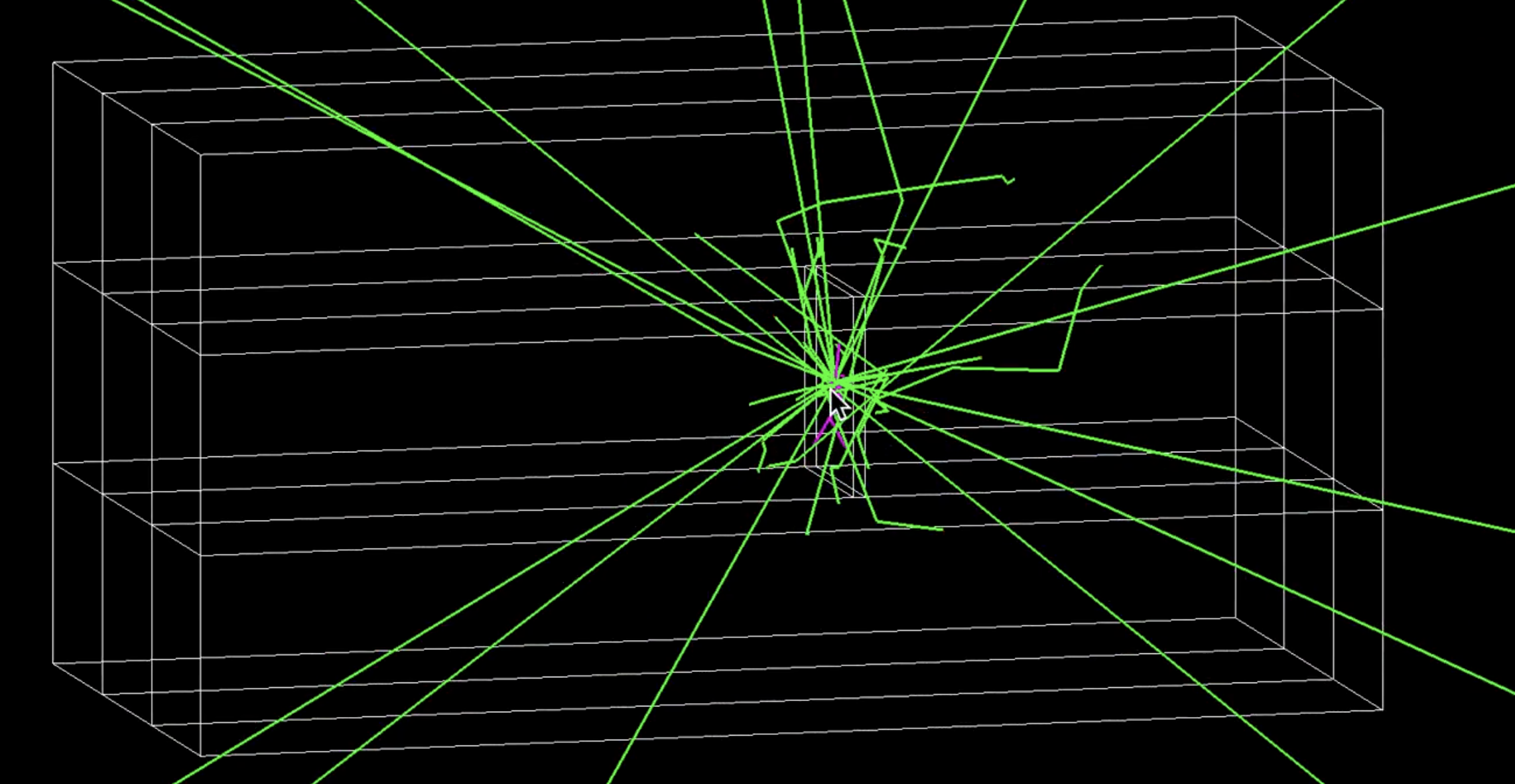}
  \caption{
  Simulation of the experimental setup. For visualization purposes, a reduced \(3\times3\) detector geometry is shown instead of the full \(5\times5\) configuration used in the analysis. The \({}^{46}\mathrm{Sc}\) radioactive source is placed at the center of the detector array, emitting $\gamma$ rays (green) and beta particles (pink). The green trajectories illustrate that a fraction of the emitted $\gamma$ rays escape the detector volume without full energy deposition, while others are absorbed within the CsI(Tl) crystals. In contrast, beta particles are rapidly absorbed near the source position because of their comparatively short range in the detector material.
  }
  \label{fig:Simulation}
\end{figure}

The detector response is modeled using the \textsc{GEANT4} simulation framework with an appropriate PhysicsList package for low-energy electromagnetic interactions. Figure~\ref{fig:Simulation} illustrates the simulated experimental setup. The simulation geometry is constructed to closely reproduce the physical detector configuration used in the experiment. The detector assembly consists of 26 CsI(Tl) crystals, including two central crystals, arranged in a \(5\times5\) square matrix. All detector elements are implemented in \textsc{Geant4} using their measured dimensions, positions, and material properties. The \({}^{46}\mathrm{Sc}\) radioactive source is placed as given in the detector, matching the experimental configuration.

For this first-order study, the simulation does not include external background sources. Consequently, passive lead shielding and the external plastic veto system are omitted from the geometry. The purpose of this simplified implementation is to estimate the intrinsic fraction of missing $\gamma$-ray events arising solely from detector containment effects for the radioactive source. Detector response effects, including finite energy resolution and threshold efficiency, are incorporated during the analysis stage by convolving the simulated energy spectra with experimentally determined detector response functions. 

\begin{figure}[h!]
\begin{subfigure}{0.48\textwidth}
  \includegraphics[width=\textwidth]{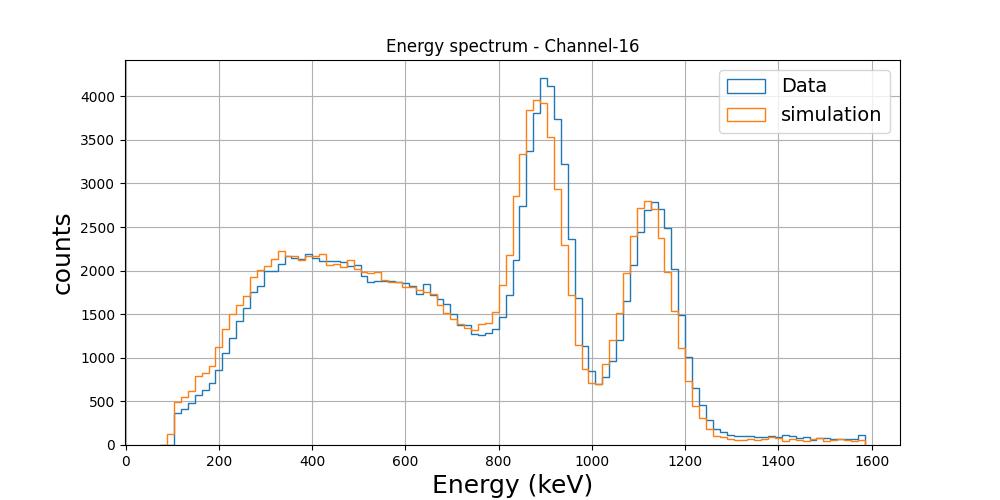}
\end{subfigure}
\caption{Energy spectrum for detector Channel 16 acquired with the ${}^{46}\mathrm{Sc}$ source. The blue histogram represents the experimentally measured spectrum, while the orange histogram shows the simulated spectrum obtained from GEANT4 after convolving with the measured detector energy resolution and threshold efficiency functions. The good agreement between data and simulation validates the detector response model built in GEANT4.}
\label{fig:data_vs_simulation_Ch16}
\end{figure}

To validate the simulation framework, simulated energy spectra were compared with experimentally measured spectra for individual detector channels. Figure~\ref{fig:data_vs_simulation_Ch16} shows the comparison for Channel 16. Figure demonstrates good agreement between simulation and experimental data after including detector response effects. In particular, the two characteristic $\gamma$-ray peaks from \({}^{46}\mathrm{Sc}\) at \(889~\mathrm{keV}\) and \(1120~\mathrm{keV}\) are well reproduced in both position and relative intensity. Similar characteristics in all the other channels also provided enough validation check for similarity between experiment and simulation. This agreement provides confidence in the accuracy of the simulation framework and supports its use for estimating detector containment efficiency and evaluating missing- $\gamma$ backgrounds.

The number of missing- $\gamma$ events is estimated by counting events within the $1120~\mathrm{keV} \pm 1.5\sigma$ region of the co-added energy spectrum across all channels. For a total exposure corresponding to $7.54 \times 10^{9}$ $^{46}$Sc decays, the simulation predicts $565.27 \pm 9.08 \times 10^{6}$ events ($7.49\%$) arising from standard detector containment losses, the cascade events in which the $889~\mathrm{keV}$  $\gamma$ ray escapes the active volume and goes undetected (Fig.~\ref{fig:Sc_spectrumAll}). This defines the baseline containment inefficiency of the detector and serves as the primary background against which the experimental missing- $\gamma$ rate is compared.

%% ---------------------------------------------------------------
%% New subsection: Backgrounds
%% To be added within Section 3 (Experimental Set-up), before Section 4
%% ---------------------------------------------------------------

\subsection{Environmental Background}
\label{sec:envbkg}

Environmental radioactivity is discussed in detail in Sec.~\ref{sec:background_and_shielding}. Using the background spectrum shown in Fig.~\ref{fig:Background}, the environmental background contribution to the missing-$\gamma$ event sample was estimated to be approximately 0.055\% by counting events within the ROI. This contribution is negligible compared to the background from incomplete $\gamma$-ray containment.

% \begin{figure}[h!]
%   \centering
%   \includegraphics[width=\columnwidth]{ASCID_Figure/Energy-AllCoin.jpg}
%   \caption{
%   Co-added energy spectrum obtained from all 26 detector channels after applying coincidence selections. The blue spectrum corresponds to the dataset without veto cuts, while the orange spectrum includes the external veto selection. The veto cut reduces the total signal rate by approximately \(5\%\).
%   }
%   \label{fig:Energy_coadded_Sc46_veto}
% \end{figure}

\subsection{Pileup as a Significant Background}
\label{sec:pileup}

\begin{figure}[h!]
\centering
  \includegraphics[width=0.9\columnwidth]{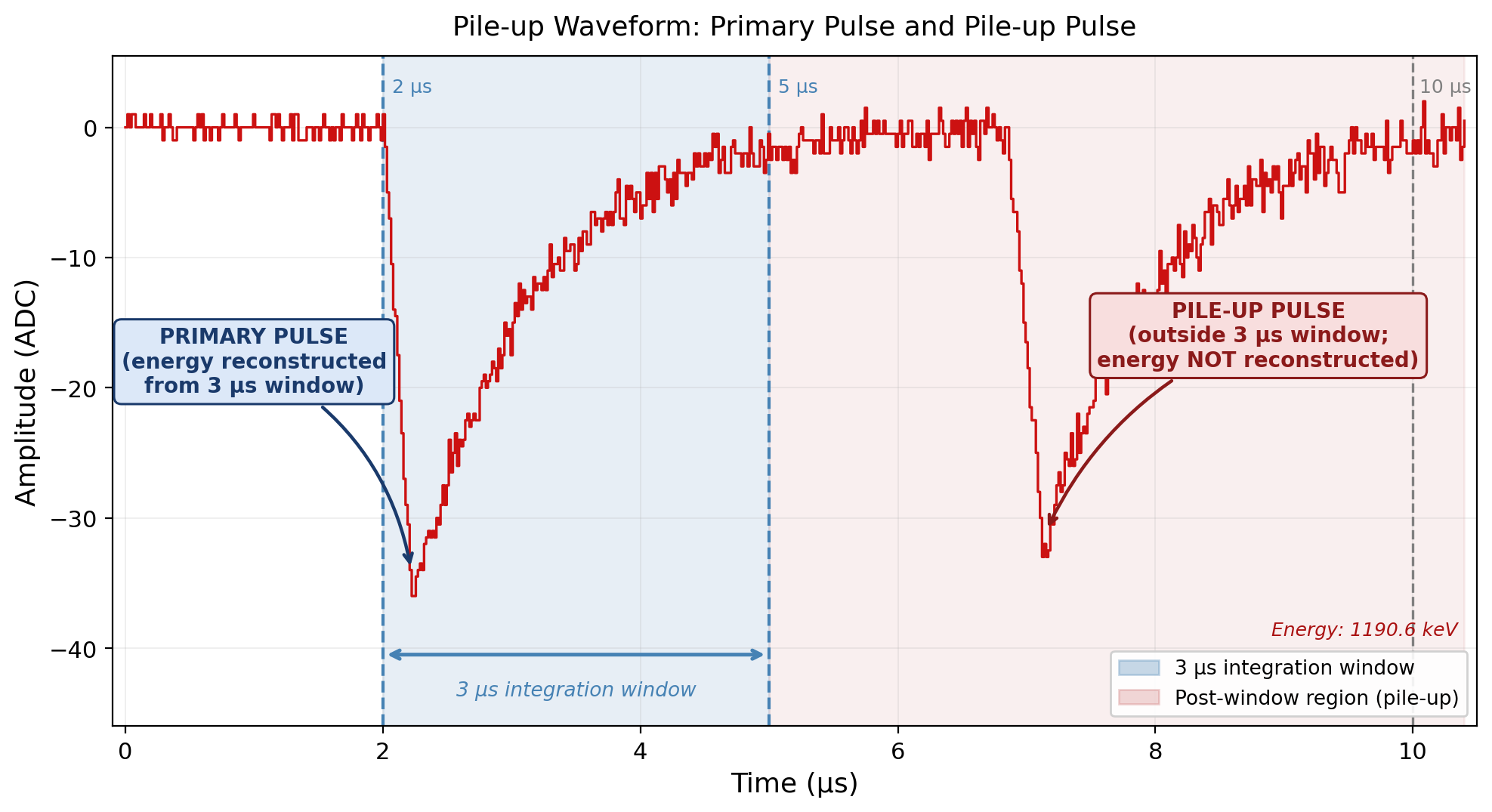}
\caption{Digitized waveform showing a pile-up event. The primary pulse is captured within the 3~$\mu$s integration window (blue); the pile-up pulse arrives in the post-window region (red) and its energy is not reconstructed, producing a false missing- $\gamma$ signature.}\label{fig:Pulse_Pileup}
\end{figure}

A significant background contribution arises from pile-up in the data acquisition system. As described earlier in Sec.~\ref{subsec:daq_system}, each recorded waveform spans a total of 10~$\mu$s, of which only a central 3~$\mu$s window is used for energy integration. If a second radioactive decay occurs within the post-pulse region of an already-triggered channel, the  $\gamma$ ray from that decay may fall partially or entirely outside the integration window and thus go unrecorded shown in Fig.~\ref{fig:Pulse_Pileup}. Such events are then misidentified as missing- $\gamma$ signatures in the subsequent trigger.

\begin{figure}[h!]
  \centering  
  \includegraphics[width=0.9\columnwidth, height = 4.6 cm]{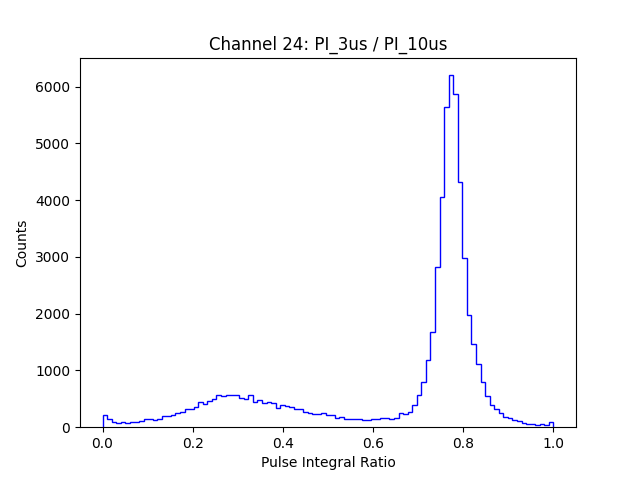}
  \caption{Distribution of the ratio between pulse integrals calculated using \(3~\mu\mathrm{s}\) and \(10~\mu\mathrm{s}\) integration windows. Two Gaussian-like populations are observed. The dominant second peak corresponds to non-pileup pulses, while the smaller first peak originates from pileup events and dark-current-induced triggers, as confirmed by inspection of representative waveforms from the respective regions of the distribution.}
  \label{fig:PI_ratio_24}
\end{figure}

The number of pile-up events $N_\text{pileup}$ is estimated by identifying pulses for which the ratio of the 3~$\mu$s integral to the full 10~$\mu$s integral falls below a characteristic threshold (Fig.~\ref{fig:PI_ratio_24}),
\begin{equation}
    \mathcal{F} = \frac{\int_0^{3\,\mu\text{s}} V(t)\,dt}{\int_0^{10\,\mu\text{s}} V(t)\,dt},
    \label{eq:pileup_filter}
\end{equation}

where $V(t)$ is the digitized pulse amplitude as a function of time. Low values of $\mathcal{F}$ indicate the presence of additional pulse energy outside the standard integration window, consistent with pile-up events. To study the impact of pile-up, a dedicated dataset was acquired in which full waveforms were recorded. This was necessary because in the high-rate data taken with high-intensity radioactive sources, storing complete waveforms for all events in the main data set was not feasible due to severe memory and data throughput limitations. The results obtained from the dedicated waveform dataset were subsequently used to model and extrapolate the pile-up contribution in the main experimental data.

In this dedicated dataset, the full waveform associated with each triggered event was recorded in addition to the reconstructed energy, detector channel, and trigger timing information. A representative subset of datasets containing full pulse information was successfully collected and analyzed to quantify the contribution of pile-up events. To identify such events, a pile-up filter was defined as the ratio of the pulse integral (PIR) calculated over a \(3~\mu\mathrm{s}\) window to that calculated over a \(10~\mu\mathrm{s}\) window, as illustrated in Fig.~\ref{fig:PI_ratio_24} for Channel 24. 

The resulting distribution exhibits two distinct gaussian populations. Events with high PIR values correspond to clean, well-contained scintillation pulses, while events with low PIR values arise primarily from either low-energy electronic noise (dark currents) or pile-up events, as confirmed through waveform inspection. Using the low filter value and an energy threshold of 100 keV, pile-up pulses were collected. To estimate the contribution of pile-up to the missing-\(\gamma\) signal region, the number of events within the energy window \(889~\mathrm{keV}\pm1.5\sigma\) was evaluated. Finally contribution for all the channels added together to estimate the exact pileup value contribution for this run. 

\begin{figure}[h!]
  \centering
  \includegraphics[width=0.9\columnwidth]{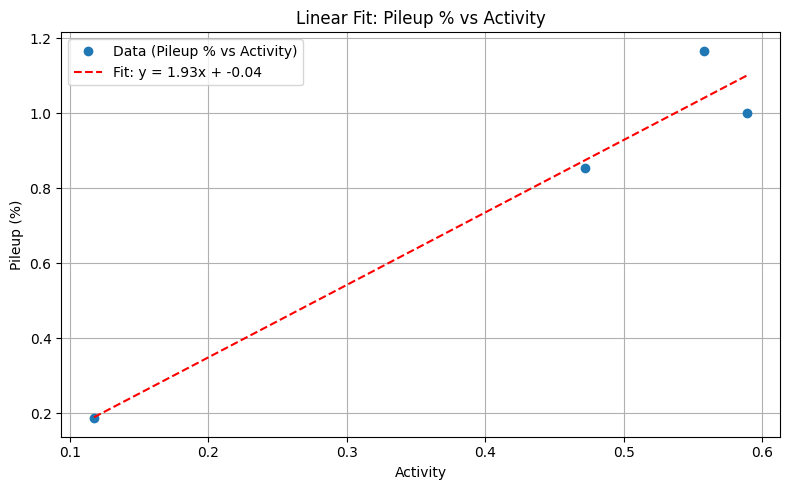}
    \caption{The plot shows the experimentally measured pileup rate for different source activities. These measurements are used to estimate the total number of pileup events accumulated over the full data-taking period.}
  \label{fig:pileup}
\end{figure}

To estimate the total number of pile-up events over the full data-taking period of the main data, the time dependence of the radioactive source activity had to be taken into account. Since the activity of the \({}^{46}\mathrm{Sc}\) source decreases with time because of radioactive decay, the pile-up rate correspondingly decreases during the experiment. To characterize this dependence, dedicated datasets were acquired using source configurations with different activity levels. From these measurements, the pile-up event rate per unit source activity was determined, as shown in Fig.~\ref{fig:pileup}.

Assuming that pile-up events occur randomly and scale linearly with source activity, this measured rate was integrated over the activity profile of the source throughout the full data-taking period to estimate the total pile-up contribution in the experimental dataset. Using this procedure, the total number of pile-up events was estimated to be \(0.95\%\) of the total events which is
\[
(70.53 \pm 0.75)\times10^{6}.
\]
After subtracting the estimated pile-up contribution, an excess of approximately
\[
49.04\times10^{6}
\]
events remained, corresponding to roughly \(0.65\%\) of the total event population. Further statistical analysis showed that this residual excess is not statistically significant in the context of the expected rare-event signal sensitivity.

\subsection{Internal Conversion: The BRICC Process}
\label{sec:bricc}

In addition, an intrinsic source of missing- $\gamma$ events is the bound-state internal conversion process, characterized by the BRICC (Band-Raman Internal Conversion Coefficients)~\cite{bricc}. In this mechanism, instead of emitting a  $\gamma$ ray, the excited nucleus transfers its de-excitation energy to an orbital electron via a virtual photon. The electron is then ejected from the atom with kinetic energy equal to the transition energy minus its binding energy. Because the emitted electron has a very short range in matter and is readily absorbed by the surrounding crystal and structural materials, it produces no detectable signal in the CsI(Tl) array. Consequently, whenever the 889~keV transition in $^{46}$Sc undergoes internal conversion rather than  $\gamma$ emission, the event appears as a genuine missing- $\gamma$ signature. Based on the BRICC tabulation~\cite{bricc}, the internal conversion coefficient for this transition is approximately 0.016\% of additional missing- $\gamma$ events to the total background budget.

\subsection{Summary of the Missing $\gamma$ Signal and the Estimated Background}

For a total exposure of $7.54 \times 10^9$ $^{46}$Sc decays, the data yield $690.19 \times 10^6$ ($9.15\%$) missing-$\gamma$ events, of which $565.27 \times 10^6$ ($7.49\%$) are attributed to detector containment inefficiency, $70.53 \times 10^6$ ($0.94\%$) to DAQ-induced pile-up, $4.16 \times 10^6$ ($0.055\%$) to environmental and radioactive backgrounds, and $1.21 \times 10^6$ ($0.016\%$) to bound-state internal conversion via the BrICC process. After subtracting all identified background contributions, a remaining excess of $49.04 \times 10^6$ ($0.65\%$) events is observed. The total systematic uncertainty is $0.373\%$, dominated by the software energy threshold ($0.267\%$) and calibration shifts ($0.21\%$), while the statistical uncertainty is negligible at $<0.001\%$. The corresponding test statistic and sensitivity plot are drawn in the following section.

\section{Conclusion And Experimental Reach}

\subsection{Exclusion Plot}
The statistical significance of the observed excess was evaluated using the test statistic~\cite{mondal2025cenunssearchcryogenicsapphire}

\begin{equation}
Z = \frac{S}{\sigma_{\text{total}}}, 
\end{equation}
where the total uncertainty is defined as 
\begin{equation}
\sigma_{\text{total}} = \sqrt{\sigma_{\text{stat}}^{2} + \sigma_{\text{sys}}^{2}}. 
\end{equation}
Using the number of decays, the estimated backgrounds, and the systematic uncertainties discussed above, the resulting significance is 

\[
\boxed{Z = 1.73}.
\]

This value falls well below the conventional thresholds typically used in rare-event searches: $Z > 5$ corresponds to a discovery claim, while $Z > 3$ is considered evidence for a possible signal. Thus, the present measurement does not indicate a statistically significant deviation attributable to new physics.

Sensitivity plot is drawn using CLs (Confidence Level) method. These are obtained by computing the exotic branching fraction for each dark-sector model using the corresponding theoretical expressions (Tab.~\ref{tab:branching_fractions}). The fundamental condition for sensitivity is that the expected signal yield equals the experimental threshold,
\begin{equation}
    \text{BR}_{\text{DM}} \times n_{\text{decay}} = \theta,
    \label{eq:sensitivity_condition}
\end{equation}
where $\theta$ depends on the statistical treatment of backgrounds. In the zero-background limit, the Poisson process gives $\theta = 2.3$ at 90\%~C.L. In the presence of background and systematics, the Gaussian approximation yields $\theta = 1.28\,\sigma_{\text{total}}$, corresponding to a 90\%~C.L.\ exclusion with mass and $g_p$, $\epsilon$, $f_{a\gamma}$ as degrees of freedom~\cite{Qian_2016}.

The sensitivity condition then becomes
\begin{equation}
    P_{\gamma \to a}(m_a,\, f_{a\gamma}) = \frac{\theta}{n_{\text{decay}}},
    \label{eq:ALP_sensitivity}
\end{equation}
where the right-hand side is determined entirely from experimental quantities. Solving Eq.~\eqref{eq:ALP_sensitivity} over the $(m_a,\, f_{a\gamma})$ parameter space exclusion plots is drawn for ALPs, Dark Scalar and Dark Photons.

\begin{figure}[h!]
\centering
\begin{minipage}{0.8\columnwidth}
    \centering
    \includegraphics[width=\linewidth]{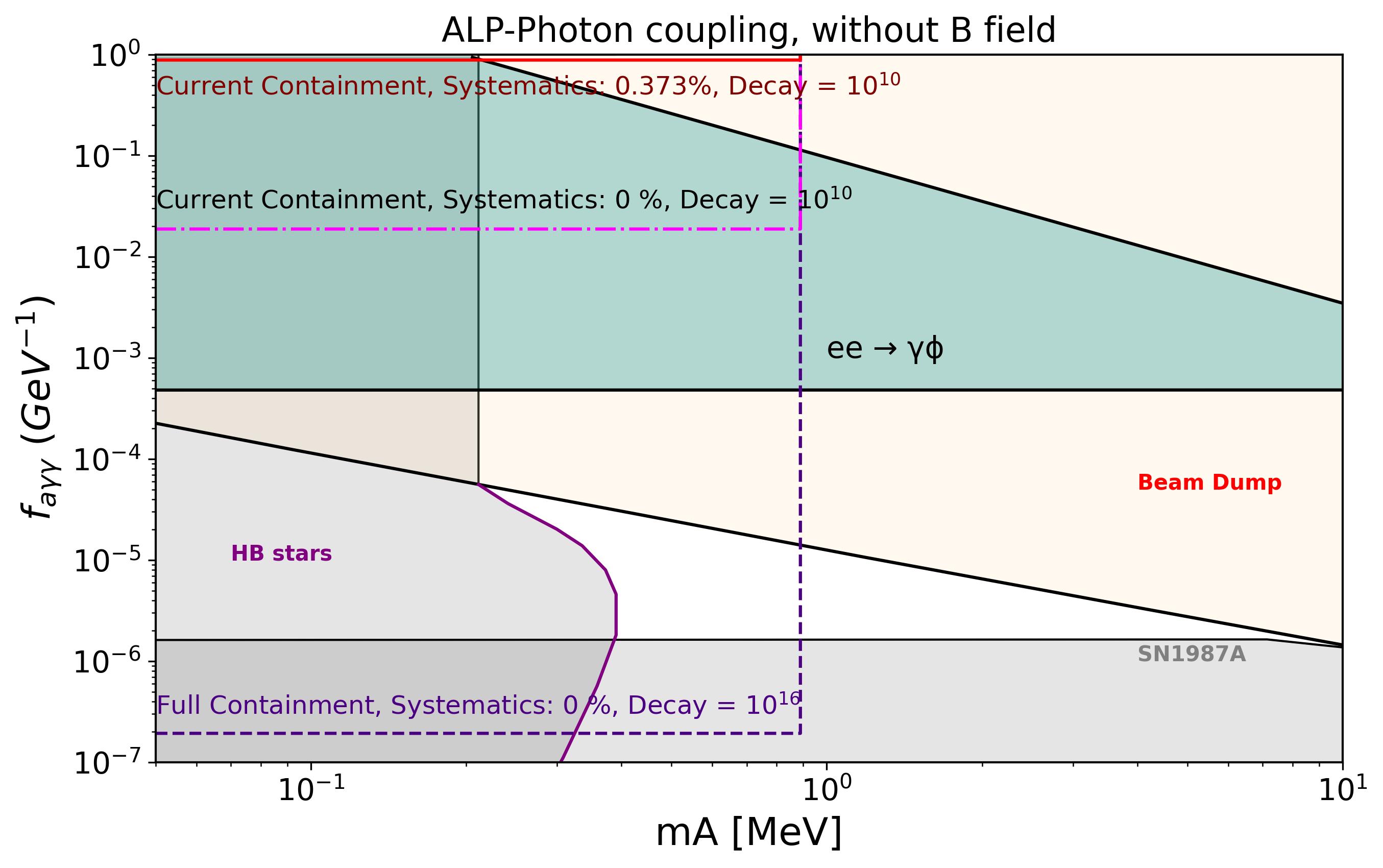}
    \caption*{(a) ALPs}
    \label{fig:fig1}
\end{minipage}

\vspace{0.3cm}

\begin{minipage}{0.8\columnwidth}
    \centering
    \includegraphics[width=\linewidth]{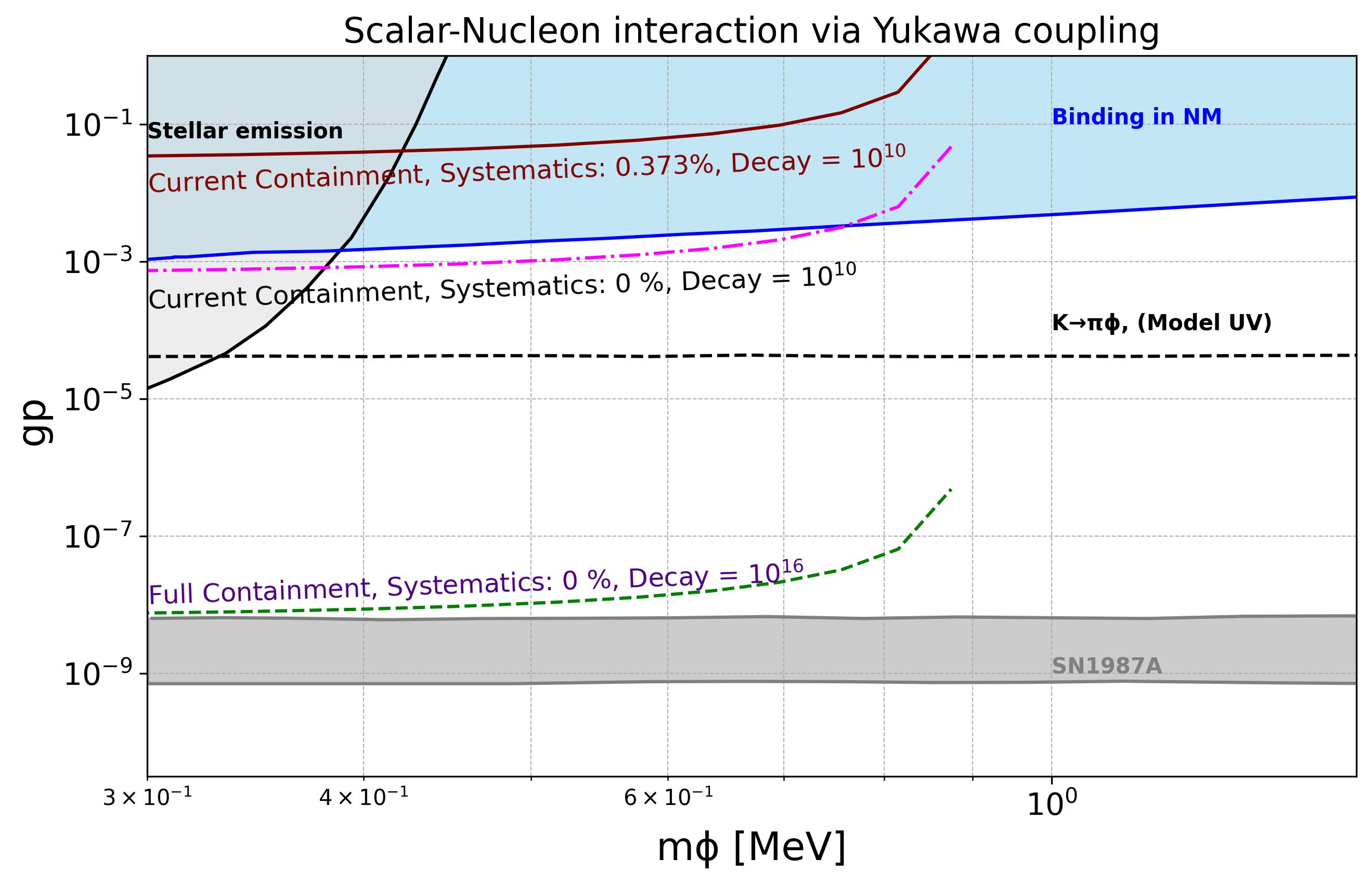}
    \caption*{(b) Scalar dark matter}
    \label{fig:fig2}
\end{minipage}

\vspace{0.3cm}

\begin{minipage}{0.8\columnwidth}
    \centering
    \includegraphics[width=\linewidth]{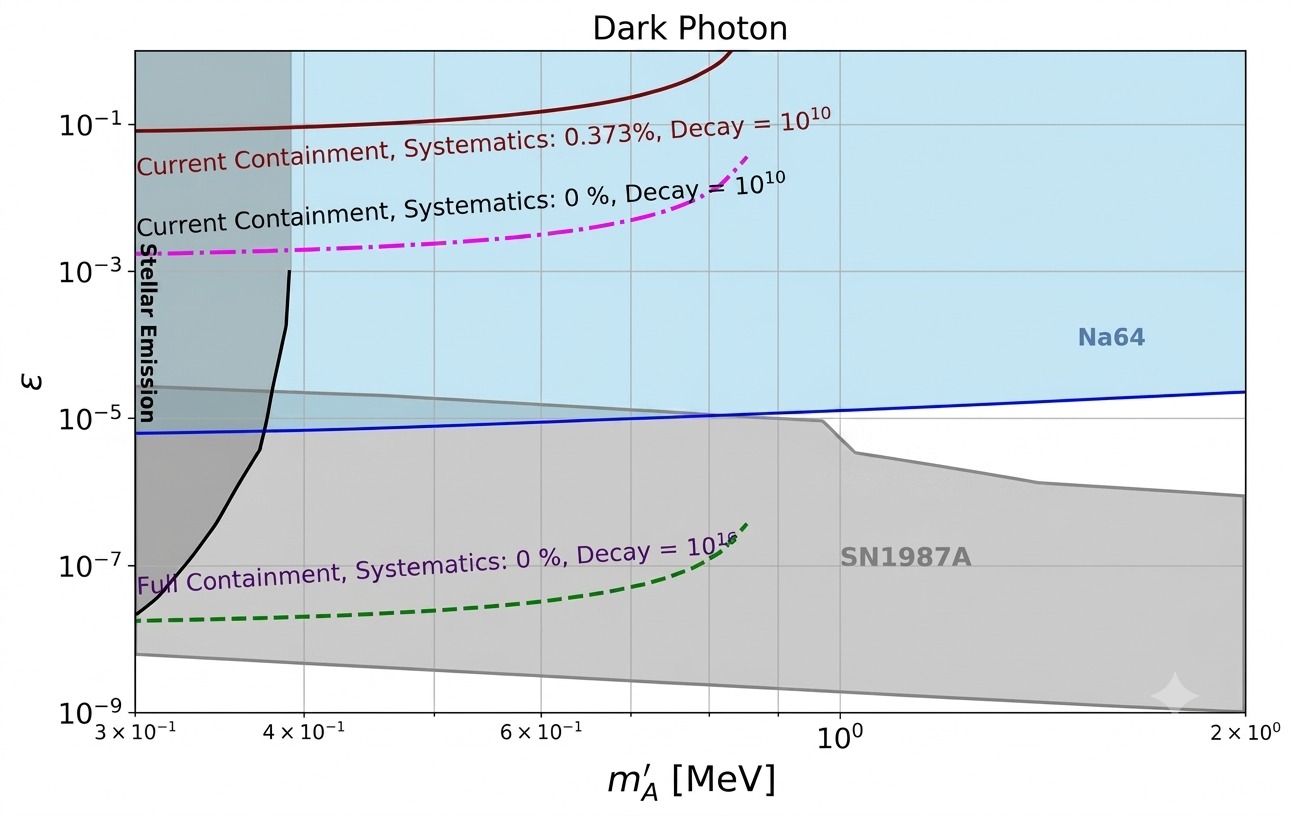}
    \caption*{(c) Dark Photon}
    \label{fig:fig3}
\end{minipage}

\caption{Exclusion limits for different dark-sector models: (a) ALPs, (b) scalar dark matter, and (c) dark photons. The solid region represents the currently excluded parameter space. The dash--dot line shows the projected sensitivity with improved systematic uncertainties, while the dashed line indicates the expected reach of a future ton-scale detector with enhanced systematics and containment efficiency. The proposed future experiments aim to probe the remaining unexplored parameter space.}\label{fig:exclusion_plots}
\end{figure}

\begin{table}[h!]
    \caption{Exploration for parameter space for different Dark-Matter candidates such as Dark Scalar, Dark Photon and ALPs.}
\centering
    \begin{tabular}{|l|c|c|c|c|}
        \hline
        Candidates & Mass & Coupling & Present& Future \\
         & (MeV) & Constant & Limit& Limit \\
        \hline
         Dark Scalar &  0.1 - 1 & $g_p$  & $> 8e-2$ &$> 1e-8$\\
         \hline
         Dark Photon & 0.1 - 1 & $\epsilon $  & $> 1e-1$&$> 1e-8$\\
        \hline
        ALP & 0.1 - 1 & $f_{a\gamma} (GeV^{-1}) $  & $> 6e-1$&$> 1e-7$\\
        \hline
    \end{tabular}
    \label{tab:DM_search}
\end{table}

The exclusion limits obtained in this study are shown in Fig.~\ref{fig:exclusion_plots}, probing regions of parameter space for axion-like particles (ALPs), scalar dark matter, and dark photons. The parameter-space ranges investigated in this work are summarized in Tab.~\ref{tab:DM_search}. Using the first-generation experiment with approximately 100~kg of CsI(Tl) scintillators, the excluded region for ALPs is relatively limited and largely overlaps with parameter space previously explored by beam-dump experiments. In the cases of scalar dark matter and dark photons, the excluded regions overlap with existing constraints from the NA64 experiment~\cite{Na64}. Nevertheless, these results validate the missing-gamma technique and demonstrate the feasibility of the experimental approach. The current sensitivity is primarily limited by detector containment, residual backgrounds, and systematic uncertainties.

Future improvements in detector mass, containment efficiency, background suppression, and systematic uncertainty control are expected to substantially enhance the experimental sensitivity, extending the reach into previously unexplored regions of parameter space and potentially probing areas that are currently constrained only by astrophysical observations.

\subsection{Future Outlook}

The primary limitations of the present experiment arise from limited $\gamma$-ray containment and relatively large systematic uncertainties, which motivate the design of the next-generation upgrade. An upgraded, ton-scale experiment is currently under development and is expected to be one of the leading lab based experiment to extend the sensitivity into previously unexplored regions of parameter space and broad, untested regions for scalar dark matter. This study highlights the importance of reducing systematic uncertainties, which are dominated by the energy threshold and the choice of coincidence window. The energy threshold is strongly influenced by the use of photomultiplier tubes (PMTs), whose gain instabilities can lead to threshold fluctuations; replacing PMTs with silicon photomultipliers (SiPMs) is expected to provide improved gain stability and enable lower, more uniform thresholds. In addition, systematic uncertainties associated with the coincidence window can be further reduced through optimized analysis strategies and faster data processing using C++-based algorithms.

Additional upgrades under consideration include the use of high-purity CsI crystals with faster decay time ($\sim 20\ ns$), which could increase statistics by up to two orders of magnitude, and the deployment of auxiliary $\beta$ detectors to improve tagging efficiency and coincidence rejection. Together, these developments will lay the foundation for the next-generation program aimed at exploring rare processes in nuclear $\gamma$ cascades with unprecedented sensitivity. Furthermore, alternative radioactive sources such as ${}^{90}\mathrm{Nb}$, ${}^{60}\mathrm{Co}$, and ${}^{24}\mathrm{Na}$ are being evaluated as substitutes for ${}^{46}\mathrm{Sc}$ to probe different dark matter mass ranges and other dark sector candidates~\cite{Missing_Gamma1}.

The detector is also being relocated to Los Alamos National Laboratory (LANL) to facilitate axion searches in a beam dump environment. A similar detector setup is planned for deployment at Oak Ridge National Laboratory (ORNL) to operate in the HFIR 85 MW reactor. This versatile detector system is designed to be compatible with a wide range of dark matter search environments, including reactor-based, beam-dump, and radioactive source experiments.

\section{Acknowledgements}

The work presented in this paper has been supported by LANL - TRIAD  and DOE Grant Nos DE-SC0018981. We also acknowledge Mitchell Institute and the Nuclear Science and Engineering Center for providing institutional support with lab space and other required facilities to carry out experimentation.

\bibliographystyle{elsarticle-num}
\bibliography{axion2}

@article{Threshold_efficiency,
author = {Angloher, G. and Banik, Sraboni and Benato, Giovanni and Bento, Antonio and Bertolini, A. and Breier, Robert and Bucci, Carlo and Burkhart, J. and Canonica, L. and D’Addabbo, A. and Conforti Di Lorenzo, Selma and Einfalt, L. and Erb, Andreas and Feilitzsch, F. v and Iachellini, N. and Fichtinger, S. and Fuchs, Dominik and Fuss, A. and Garai, A. and Zema, V.},
year = {2023},
month = {06},
pages = {},
title = {Results on sub-GeV dark matter from a 10 eV threshold CRESST-III silicon detector},
volume = {107},
journal = {Physical Review D},
doi = {10.1103/PhysRevD.107.122003}
}

@misc{GammaData2024_CsITlNa_DataSheet,
  author       = {{GammaData}},
  title        = {CsI(Tl) and Na Scintillation Material Data Sheet},
  year         = {2024},
  howpublished = {\url{https://gammadata.se/wp-content/uploads/2024/01/CsITl-and-Na-data-sheet-v2.pdf}},
  note         = {Technical datasheet},
}

@misc{PMT_response,
  author = {{ET Enterprises}},
  title  = {{PMT 9954B Response Curve}},
  year   = {2025},
  note   = {Available at: \url{https://et-enterprises.com/images/data_sheets/9954B.pdf} (Accessed: 2025-02-18)}
}

@misc{PTFE_Tape,
  author = {{Thor Labs}},
  title  = {{High-Reflectance PTFE Sheets
}},
  year   = {2020},
  note   = {Available at: \url{https://www.thorlabs.com/high-reflectance-ptfe-sheets?tabName=Overview
}}
}

@misc{HV_SY5527LC,
  author = {{CAEN HV SY5527LC}},
  title  = {{SY5527LC
Universal Multichannel Power Supply System (Low Cost)}},
  year   = {2026},
  note   = {Available at: \url{https://www.caen.it/products/sy5527lc/
}}
}

@misc{HV_A7236,
  author = {{CAEN HV A7236}},
  title  = {{12/24/32 Channel 3.5 kV, 1.5/0.15 mA (4W) Common Floating Return Dual Range Boards}},
  year   = {2026},
  note   = {Available at: \url{https://www.caen.it/products/a7236/
}}
}

@misc{CAEN_V1740D,
  author = {{CAEN Digitizer V1740D}},
  title  = {{64 Channel 12 bit 62.5 MS/s Digitizer supporting DPP-QDC firmware}},
  year   = {2026},
  note   = {Available at: \url{https://www.caen.it/products/v1740d/
}}}

@misc{GECO2020,
  author = {{CAEN GECO2020}},
  title  = {{GEneral COntrol Software for CAEN HV Power Supplies}},
  year   = {2026},
  note   = {Available at: \url{https://www.caen.it/products/geco2020/
}}}

@misc{COMPASS,
  author = {{CAEN COMPASS}},
  title  = {{Multiparametric DAQ Software for Physics Applications}},
  year   = {2026},
  note   = {Available at: \url{https://www.caen.it/products/compass/
}}}

@misc{Al_Myler,
  author = {{3M™ }},
  title  = {{3M™ EMI Aluminum Foil Shielding Tape 1170
}},
  year   = {2021},
  note   = {Available at: \url{https://www.3m.com/3M/en_US/p/d/b00041317/}}
}

@misc{Lead_Shielding,
  author = {{FIRGELLI Automations}},
  title  = {{Shielding Thickness Gamma Interactive Calculator}},
  year   = {2026},
  note   = {Available at: \url{https://www.firgelliauto.com/blogs/calculators/shielding-thickness-gamma-calculator?srsltid=AfmBOorGhv_wVYU4ajqlg9Xp7D1czA3J3YV-BmGr9zisOzCaddPpOuSF}}
}

@misc{Plastic_Scintillator,
  author = {{Luxium Solutions}},
  title  = {{BC-400, BC-404, BC-408, BC-412, BC-416}},
  year   = {2026},
  note   = {Available at: \url{https://luxiumsolutions.com/radiation-detection-scintillators/plastic-scintillators/bc400-bc404-bc408-bc412-bc416
}}}

@article{Aprile_2020,
   title={Energy resolution and linearity of XENON1T in the MeV energy range},
   volume={80},
   ISSN={1434-6052},
   url={http://dx.doi.org/10.1140/epjc/s10052-020-8284-0},
   DOI={10.1140/epjc/s10052-020-8284-0},
   number={8},
   journal={The European Physical Journal C},
   publisher={Springer Science and Business Media LLC},
   author={Aprile, E. and Aalbers et al.},
   year={2020},
   month=Aug }

@article{Energy_Linearity,
author = {Yavuzkanat, Nuray and Şenyiğit, M. and Kaplan, Muhammet},
year = {2022},
month = {06},
pages = {},
title = {Investigation of the gamma-ray efficiency for various scintillation detector systems},
volume = {68},
journal = {Revista Mexicana de Física},
doi = {10.31349/RevMexFis.68.041201}
}

@misc{Na64,
      title={Searching for Light Dark Matter and Dark Sectors with the NA64 experiment at the CERN SPS}, 
      author={Yu. M. Andreev and A. Antonov and M. A. Ayala Torres and D. Banerjee and B. Banto Oberhauser and V. Bautin and J. Bernhard and P. Bisio and A. Celentano and N. Charitonidis and P. Crivelli and A. V. Dermenev and S. V. Donskov and R. R. Dusaev and V. N. Frolov and S. V. Gertsenberger and S. Girod and S. N. Gninenko and A. V. Ivanov and Y. Kambar and A. E. Karneyeu and G. Kekelidze and B. Ketzer and D. V. Kirpichnikov and M. M. Kirsanov and V. A. Kramarenko and N. V. Krasnikov and S. V. Kuleshov and V. E. Lyubovitskij and A. Marini and L. Marsicano and V. A. Matveev and R. Mena Fredes and R. Mena Yanssen and L. Molina Bueno and M. Mongillo and D. V. Peshekhonov and V. A. Polyakov and B. Radics and K. Salamatin and V. D. Samoylenko and H. Sieber and D. Shchukin and O. Soto and V. O. Tikhomirov and I. Tlisova and A. N. Toropin and M. Tuzi and P. V. Volkov and I. V. Voronchikhin and J. Zamora-Saá and A. S. Zhevlakov},
      year={2025},
      eprint={2505.14291},
      archivePrefix={arXiv},
      primaryClass={hep-ex},
      url={https://arxiv.org/abs/2505.14291}, 
}

@misc{statistical_uncertainty,
      title={Confidence intervals for the Poisson distribution}, 
      author={Frank C. Porter},
      year={2026},
      eprint={2509.02852},
      archivePrefix={arXiv},
      primaryClass={physics.data-an},
      url={https://arxiv.org/abs/2509.02852}, 
}

@misc{systematic_uncertainty,
      title={Definition and Treatment of Systematic Uncertainties in High Energy Physics and Astrophysics}, 
      author={Pekka K. Sinervo and C. M},
      year={2025},
      eprint={2510.24313},
      archivePrefix={arXiv},
      primaryClass={hep-ex},
      url={https://arxiv.org/abs/2510.24313}, 
}

@article{LI2024, 
author = {Gang LI and Yaoqi WANG and Xiaopeng WANG},
title = {Adaptive compensation method for photomultiplier tube counting temperature drift},
year = {2024},
journal = {Journal of Measurement Science and Instrumentation},
volume = {15},
number = {2},
pages = {244-252},
url = {https://www.sciopen.com/article/10.62756/jmsi.1674-8042.2024025},
doi = {10.62756/jmsi.1674-8042.2024025}
}

@article{PhysRevLett.106.131302,
  title = {Results from a Low-Energy Analysis of the CDMS II Germanium Data},
  author = {Ahmed, Z. et.al.},
  collaboration = {CDMS Collaboration},
  journal = {Phys. Rev. Lett.},
  volume = {106},
  issue = {13},
  pages = {131302},
  numpages = {5},
  year = {2011},
  month = {Mar},
  publisher = {American Physical Society},
  doi = {10.1103/PhysRevLett.106.131302},
  url = {https://link.aps.org/doi/10.1103/PhysRevLett.106.131302}
}

@misc{mondal2025cenunssearchcryogenicsapphire,
      title={CE$\nu$NS Search with Cryogenic Sapphire Detectors at MINER: Results from the TRIGA reactor data and Future Sensitivity at HFIR}, 
      author={D. Mondal and W. Baker and M. Chaudhuri and J. B. Dent and R. Dey and B. Dutta and V. Iyer and A. Jastram and V. K. S. Kashyap and A. Kubik and K. Lang and R. Mahapatra and S. Maludze and N. Mirabolfathi and M. Mirzakhani and B. Mohanty and H. Neog and J. L. Newstead and M. Platt and S. Sahoo and J. Sander and L. E. Strigari and J. Walker},
      year={2025},
      eprint={2510.15999},
      archivePrefix={arXiv},
      primaryClass={nucl-ex},
      url={https://arxiv.org/abs/2510.15999}, 
}

@article{Qian_2016,
   title={The Gaussian CL  method for searches of new physics},
   volume={827},
   ISSN={0168-9002},
   url={http://dx.doi.org/10.1016/j.nima.2016.04.089},
   DOI={10.1016/j.nima.2016.04.089},
   journal={Nuclear Instruments and Methods in Physics Research Section A: Accelerators, Spectrometers, Detectors and Associated Equipment},
   publisher={Elsevier BV},
   author={Qian, X. and Tan, A. and Ling, J.J. and Nakajima, Y. and Zhang, C.},
   year={2016},
   month=Aug, pages={63–78} }

@misc{BRICC,
  author = {{Australian National University}},
  title  = {BRICC Conversion Coefficient Database},
  year   = {2025},
  note   = {Available at: \url{https://bricc.anu.edu.au/} (Accessed: October 2025)}
}

@article{Baur_2019,
   title={Correction of beam hardening in X-ray radiograms},
   volume={90},
   ISSN={1089-7623},
   url={http://dx.doi.org/10.1063/1.5080540},
   DOI={10.1063/1.5080540},
   number={2},
   journal={Review of Scientific Instruments},
   publisher={AIP Publishing},
   author={Baur, Manuel and Uhlmann, Norman and Pöschel, Thorsten and Schröter, Matthias},
   year={2019},
   month=Feb }

@misc{NIST_XCOM,
  author       = {Berger, M.J. and Hubbell, J.H. and Seltzer, S.M. and Chang, J. 
                  and Coursey, J.S. and Sukumar, R. and Zucker, D.S. and Olsen, K.},
  title        = {{XCOM}: Photon Cross Sections Database},
  howpublished = {NIST Standard Reference Database 8 (XGAM), 
                  National Institute of Standards and Technology},
  year         = {2010},
  note         = {Available at \url{https://physics.nist.gov/PhysRefData/Xcom/html/xcom1.html}},
}

@misc{sahoo2026reactorbasedsearchaxionlikeparticles,
      title={Reactor-based Search for Axion-Like Particles using CsI(Tl) Detector}, 
      author={S. Sahoo and S. Verma and M. Mirzakhani and N. Mishra and A. Thompson and S. Maludze and R. Mahapatra and M. Platt},
      year={2026},
      eprint={2407.14704},
      archivePrefix={arXiv},
      primaryClass={hep-ex},
      url={https://arxiv.org/abs/2407.14704}, 
}

@techreport{OpticalGlue_response,
  author      = {{Eljen Technology}},
  institution = {Eljen Technology},
  title       = {{EJ-550 and EJ-552 Optical Coupling Compounds: Technical Datasheet}},
  year        = {2020},
  type        = {Technical report},
  url         = {https://eljentechnology.com/products/accessories/ej-550-ej-552},
  note        = {Accessed: 2025-02-18}
}

@phdthesis{Verma,
    author = {Verma, Shubham Dilip},
    title = {Fabrication of CDMS Dark Matter Detector Using Bi-layer Lift-off Technique and Detection of Axions/Axion Like Particles (Alps) Using CsI(Tl) Scintillator Detectors},
    school = {Texas A-M},
    year = {2022}
}

@article{VINCENT2023168624,
title = {Improving the resolution and light yield in CsI(Tl) scintillators},
journal = {Nuclear Instruments and Methods in Physics Research Section A: Accelerators, Spectrometers, Detectors and Associated Equipment},
volume = {1056},
pages = {168624},
year = {2023},
issn = {0168-9002},
doi = {https://doi.org/10.1016/j.nima.2023.168624},
url = {https://www.sciencedirect.com/science/article/pii/S0168900223006149},
author = {Kensington N. Vincent and Samikshya Mahapatra and Ishita Poddar and Shubham Verma},

}

@book{Cherry2012_NuclearMedicine,
  author    = {Cherry, S. R. and Sorenson, J. A. and Phelps, M. E.},
  title     = {Physics in Nuclear Medicine},
  edition   = {4},
  publisher = {Elsevier},
  year      = {2012}
}

@article{CLEO,
    author = "Blucher, E. and Gittelman, B. and Heltsley, B. K. and Kandaswamy, J. and Kowalewski, Robert V. and Kubota, Y. and Mistry, Nari B. and Stone, S. and Bean, A.",
    title = "{Tests of Cesium Iodide Crystals for an Electromagnetic Calorimeter}",
    reportNumber = "CLNS-86-721",
    doi = "10.1016/0168-9002(86)90669-8",
    journal = "Nucl. Instrum. Meth. A",
    volume = "249",
    pages = "201",
    year = "1986"
}

@article{Missing_Gamma2,
   title={Pathfinder for a high statistics search for missing energy in gamma cascades},
   volume={105},
   ISSN={2470-0029},
   url={http://dx.doi.org/10.1103/PhysRevD.105.015030},
   DOI={10.1103/physrevd.105.015030},
   number={1},
   journal={Physical Review D},
   publisher={American Physical Society (APS)},
   author={Dent, James B. and Dutta, Bhaskar and Jastram, Andrew and Kim, Doojin and Kubik, Andrew and Mahapatra, Rupak and Rajendran, Surjeet and Ramani, Harikrishnan and Thompson, Adrian and Verma, Shubham},
   year={2022},
   month=jan }

@article{Missing_Gamma1,
  title = {Invisible decay modes in nuclear gamma cascades},
  author = {Benato, Giovanni and Drobizhev, Alexey and Rajendran, Surjeet and Ramani, Harikrishnan},
  journal = {Phys. Rev. D},
  volume = {99},
  issue = {3},
  pages = {035025},
  numpages = {10},
  year = {2019},
  month = {Feb},
  publisher = {American Physical Society},
  doi = {10.1103/PhysRevD.99.035025},
  url = {https://link.aps.org/doi/10.1103/PhysRevD.99.035025}
}

@article{alcock,
  title={Possible gravitational microlensing of a star in the Large Magellanic Cloud},
  author={Alcock, Charles and Akerlof, Carl W and Allsman, RA and Axelrod, TS and Bennett, DP and Chan, S and Cook, KH and Freeman, Kenneth C and Griest, K and Marshall, Stuart L and others},
  journal={nature},
  volume={365},
  number={6447},
  pages={621--623},
  year={1993},
  publisher={Nature Publishing Group UK London}
}

@article{Zwicky1,
    author = "Zwicky, F.",
    title = "{Die Rotverschiebung von extragalaktischen Nebeln}",
    doi = "10.1007/s10714-008-0707-4",
    journal = "Helv. Phys. Acta",
    volume = "6",
    pages = "110--127",
    year = "1933"
}

@inbook{Zwicky2,
url = {https://doi.org/10.4159/harvard.9780674366688.c115},
title = {107. On the Masses of Nebulae and of Clusters of Nebulae},
booktitle = {A Source Book in Astronomy and Astrophysics, 1900–1975},
author = {Fritz Zwicky},
editor = {Kenneth R. Lang and Owen Gingerich},
publisher = {Harvard University Press},
address = {Cambridge, MA and London, England},
pages = {729--737},
doi = {doi:10.4159/harvard.9780674366688.c115},
isbn = {9780674366688},
year = {1979},
lastchecked = {2024-05-27}
}

@article{Mori:2021pcv,
    author = "Mori, Kanji and Takiwaki, Tomoya and Kotake, Kei and Horiuchi, Shunsaku",
    title = "{Shock revival in core-collapse supernovae assisted by heavy axionlike particles}",
    eprint = "2112.03613",
    archivePrefix = "arXiv",
    primaryClass = "astro-ph.HE",
    doi = "10.1103/PhysRevD.105.063009",
    journal = "Phys. Rev. D",
    volume = "105",
    number = "6",
    pages = "063009",
    year = "2022"
}

@article{Graham2019,
   title={Relaxation of the cosmological constant},
   volume={100},
   ISSN={2470-0029},
   url={http://dx.doi.org/10.1103/PhysRevD.100.015048},
   DOI={10.1103/physrevd.100.015048},
   number={1},
   journal={Physical Review D},
   publisher={American Physical Society (APS)},
   author={Graham, Peter W. and Kaplan, David E. and Rajendran, Surjeet},
   year={2019},
   month=jul }

@article{Graham_2015,
   title={Cosmological Relaxation of the Electroweak Scale},
   volume={115},
   ISSN={1079-7114},
   url={http://dx.doi.org/10.1103/PhysRevLett.115.221801},
   DOI={10.1103/physrevlett.115.221801},
   number={22},
   journal={Physical Review Letters},
   publisher={American Physical Society (APS)},
   author={Graham, Peter W. and Kaplan, David E. and Rajendran, Surjeet},
   year={2015},
   month=nov }

@article{Wilczek1978,
  title = {Problem of Strong $P$ and $T$ Invariance in the Presence of Instantons},
  author = {Wilczek, F.},
  journal = {Phys. Rev. Lett.},
  volume = {40},
  issue = {5},
  pages = {279--282},
  numpages = {0},
  year = {1978},
  month = {Jan},
  publisher = {American Physical Society},
  doi = {10.1103/PhysRevLett.40.279},
  url = {https://link.aps.org/doi/10.1103/PhysRevLett.40.279}
}

@article{Weinberg1978,
  title = {A New Light Boson?},
  author = {Weinberg, Steven},
  journal = {Phys. Rev. Lett.},
  volume = {40},
  issue = {4},
  pages = {223--226},
  numpages = {0},
  year = {1978},
  month = {Jan},
  publisher = {American Physical Society},
  doi = {10.1103/PhysRevLett.40.223},
  url = {https://link.aps.org/doi/10.1103/PhysRevLett.40.223}
}

@article{PhysRevD.16.1791,
  title = {Constraints imposed by $\mathrm{CP}$ conservation in the presence of pseudoparticles},
  author = {Peccei, R. D. and Quinn, Helen R.},
  journal = {Phys. Rev. D},
  volume = {16},
  issue = {6},
  pages = {1791--1797},
  numpages = {0},
  year = {1977},
  month = {Sep},
  publisher = {American Physical Society},
  doi = {10.1103/PhysRevD.16.1791},
  url = {https://link.aps.org/doi/10.1103/PhysRevD.16.1791}
}

@article{Rubin1,
  title   = "Rotation of the Andromeda Nebula from a
Spectroscopic Survey of Emission Regions",
  journal = "Astrophys. J.",
  volume  = "159",
  number  = "",
  pages   = "379-403",
  year    = "1970",
  doi     = "",
  author  = "V. C. Rubin and W. K. Ford, Jr."
}

\section{Appendices}
\label{sec:Appendices}

\subsection{Missing $\gamma$ Fraction Calculation}
\label{app:missing_containment_analytical}

The missing fraction $f_{\rm miss}$ for the 889~keV $\gamma$ ray from $^{46}$Sc was computed as follows. The total mass attenuation coefficient of CsI(Tl) at 889~keV, excluding coherent (Rayleigh) scattering which deposits negligible energy, is $\mu/\rho = 6.132\times10^{-2}$~cm$^{2}$/g~\cite{NIST_XCOM}. Combined with the density $\rho = 4.51$~g/cm$^{3}$, this gives a linear attenuation coefficient $\mu = 0.2766$~cm$^{-1}$ and mean free path $\lambda = 3.62$~cm. The $^{46}$Sc point source is located at the origin of a rectangular CsI(Tl) array with face distances of $\pm$10.16~cm in $X$ and $Y$, $+$18.29~cm in $+Z$, and $-$12.19~cm in $-Z$. For an isotropically emitted $\gamma$ in direction $(\theta,\phi)$, the path length to the nearest crystal face is
\begin{equation}
    d_{\min}(\theta,\phi) = \min\!\left(
    \frac{10.16}{|\sin\theta\cos\phi|},\;
    \frac{10.16}{|\sin\theta\sin\phi|},\;
    \frac{z_{\rm face}}{|\cos\theta|}
    \right),
    \label{eq:dmin}
\end{equation}
where $z_{\rm face} = 18.29$~cm for $\cos\theta > 0$ and $12.19$~cm otherwise. The escape probability for that direction is $e^{-\mu d_{\min}}$, and the total missing fraction is the $4\pi$-averaged escape probability,
\begin{equation}
    f_{\rm miss} = \frac{1}{4\pi}\int_{0}^{2\pi}d\phi
    \int_{0}^{\pi}\sin\theta\;e^{-\mu\,d_{\min}(\theta,\phi)}\;d\theta.
    \label{eq:fmiss}
\end{equation}
This integral was evaluated numerically by Monte Carlo sampling of $2\times10^{7}$ uniformly distributed directions on the unit sphere, yielding $f_{\rm miss} = (2.773\pm0.004)\%$. This represents a geometric lower bound; the ($-Z$, 12.19~cm) face corresponds to ${\sim}3.4$ mean free paths and the lateral faces ($\pm X$, $\pm Y$, 10.16~cm) to ${\sim}2.8$ mean free paths, so exponential attenuation strongly suppresses escape in all directions.

\subsection{Estimation of Time Coincidence Window}
\label{sec:coincidence_window}

To associate all signals originating from a single physical decay, a coincidence window is defined in the analysis. Events occurring within this window are grouped together and treated as a single decay event, ensuring that correlated multi-hit signals are not mistakenly identified as independent events.

\begin{figure}[h!]
\centering
\begin{subfigure}{0.33\textwidth}
  \includegraphics[width=\linewidth]{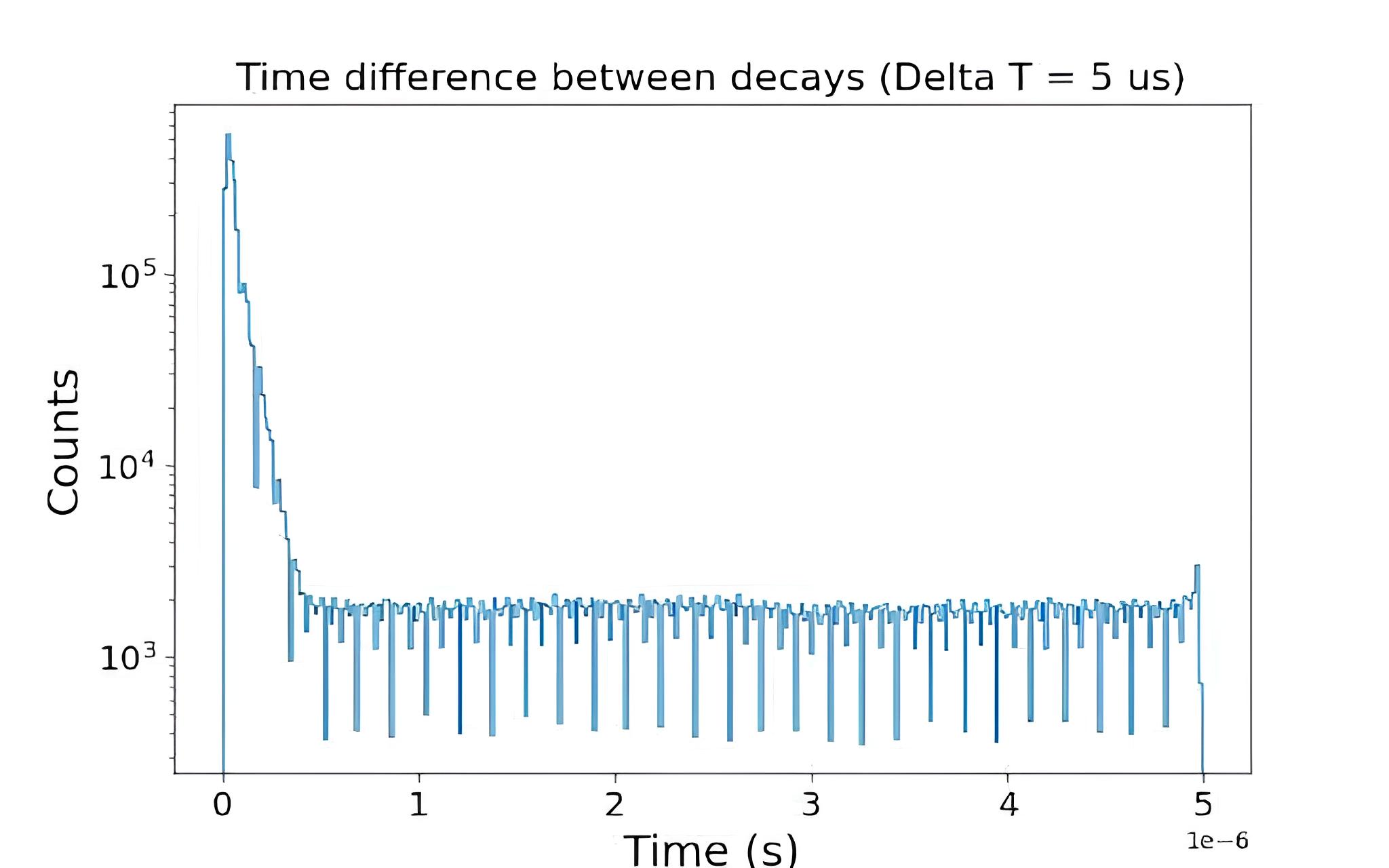}
  \caption{Log plot.}
\end{subfigure}%
\hfill
\begin{subfigure}{0.14\textwidth}
  \includegraphics[width=\linewidth]{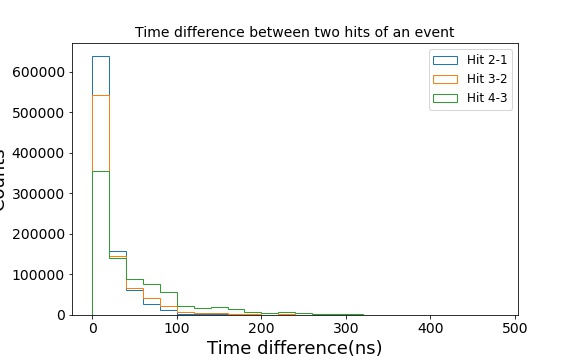}

  \vspace{0.3cm}

  \includegraphics[width=\linewidth]{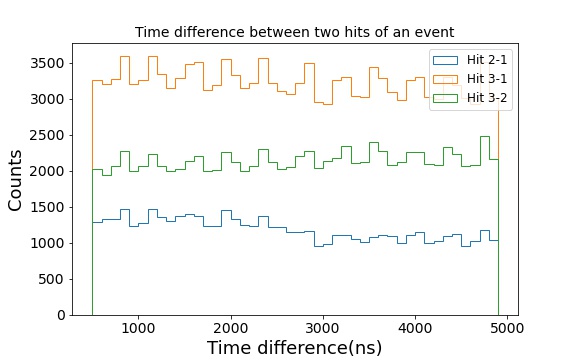}
  \caption{Linear.}
\end{subfigure}
\caption[Histogram of time differences between consecutive events.]{Histogram of the time differences between consecutive events. A pronounced exponential component is observed for time intervals below $0.5\,\mu\mathrm{s}$, corresponding to correlated radioactive decay events, while the approximately flat distribution at larger time differences is consistent with uncorrelated background events. Panel (a) shows the event rate on a semilogarithmic scale for time differences between 0 and $5\,\mu\mathrm{s}$. Panel (b) presents the same distribution on a linear scale for $0$--$0.5\,\mu\mathrm{s}$ (top) and $0.5$--$5\,\mu\mathrm{s}$ (bottom).}
\label{fig:Coincidence}
\end{figure}

The distribution of time differences between successive hits exhibits a pronounced excess at short time separations, indicating correlated signals originating from the same decay, while larger separations are consistent with uncorrelated background events.

Based on the observed event clustering below 0.5~$\mu$s (Fig.~\ref{fig:Coincidence}), the 16~ns digitizer time resolution, the $\sim$3~$\mu$s CsI(Tl) decay time constant, and the measured pulse rise times, a coincidence window of 0.5~$\mu$s was adopted for event grouping. Any loss of correlated events outside this window was treated as a systematic uncertainty and propagated through the final analysis.

\subsection{Preliminary Analysis with ${}^{22}\mathrm{Na}$}
\label{app:Na22_analysis}

\begin{figure}[h!]
\centering

\begin{subfigure}[b]{0.48\columnwidth}
    \centering
    \includegraphics[width=\columnwidth]{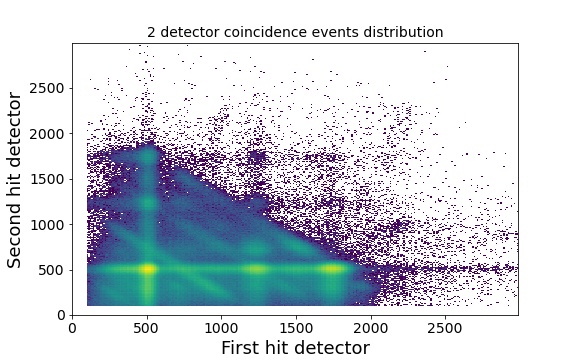}
    \caption{Experimental data.}
    \label{fig:2det_data}
\end{subfigure}
\hfill
\begin{subfigure}[b]{0.48\columnwidth}
    \centering
    \includegraphics[width=\columnwidth]{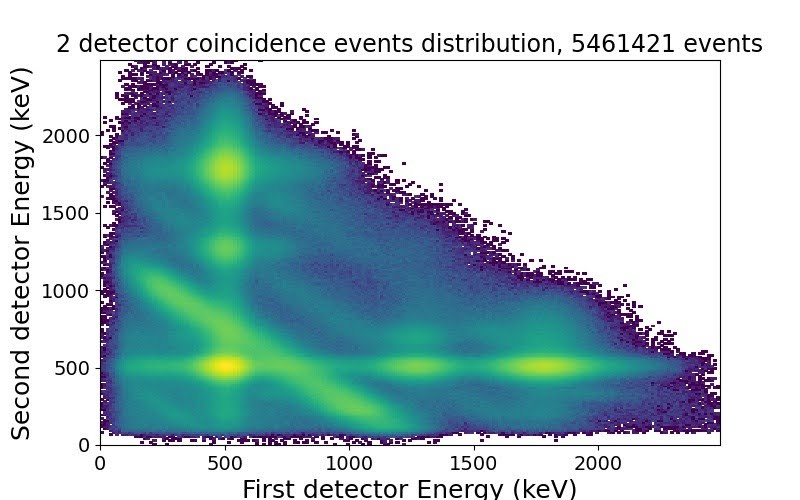}
    \caption{Monte Carlo simulation.}
    \label{fig:2det_sim}
\end{subfigure}

\caption{Two-detector shared-energy spectra for coincident events in \({}^{22}\mathrm{Na}\) data (a) and Monte Carlo simulation (b). Distinct clusters corresponding to the expected $\gamma$-ray energy combinations and diagonal bands arising from energy sharing between detectors are reproduced by the simulation, demonstrating good agreement with the measured data and validating the detector response model.}
\label{fig:2det_3_3}

\end{figure}

While a comprehensive analysis of the \({}^{22}\mathrm{Na}\)  data is presented in Sec.~\ref{sec:Na22:analysis}, an additional comparison is included here to further demonstrate detector performance and validate the Monte Carlo simulation.

A detailed examination was performed for events within 2-detector multiplicity category, defined as cases in which exactly two crystals triggered within the coincidence window. Figure~\ref{fig:2det_3_3} shows a two-dimensional histogram of the energy deposited in the first triggered detector versus that in the second triggered detector. Distinct clusters are observed at the expected energy combinations (511, 511), (511, 1275), (511, 1786), (1275, 511), and (1786, 511)~keV, corresponding to the possible $\gamma$-ray combinations produced in ${}^{22}\mathrm{Na}$ decays.

In addition to these clusters, diagonal bands are visible at total energies of 1275~keV, 1786~keV, and 2297~keV, indicating events in which a $\gamma$ ray shares its energy between two detectors, most likely through Compton scattering or partial energy deposition. A slight asymmetry is also observed in the distribution, with brighter clusters associated with the first triggered detector. This effect arises because higher-energy $\gamma$ rays, such as those at 1275~keV or 1786~keV, tend to produce faster-rising pulses that cross the digitizer threshold earlier, making them more likely to be recorded as the first trigger in the detector pair.

\subsection{Errors and Uncertainty}
\label{sec:uncertainty}

A summary of the statistical and systematic uncertainties is provided in Sec.~\ref{sec:statistical_systematic}, while a detailed discussion of their evaluation is presented in this section. In general, the uncertainties associated with the experiment can be broadly classified into two categories:

\begin{itemize}
    \item \textbf{Statistical uncertainty:} This quantifies the uncertainty arising from finite data samples and random fluctuations, and it decreases as the amount of collected data increases.
    \item \textbf{Systematic uncertainty:} This arises from potential biases in the experimental setup and analysis procedures.
\end{itemize}

\subsubsection{Calculation of Statistical Uncertainty}

The statistical uncertainty is straightforward to estimate. For the large data sets considered in this analysis, the statistical fluctuations are well described by a Gaussian distribution and given by the square root of the number of events in each case \cite{statistical_uncertainty}.

\begin{align*}
    \sigma_{\text{stat}} &= \sqrt{\sum_i \sigma_{\text{stat},B_i}^2} \\
    \sigma_{\text{stat},B_i} &= \sqrt{N_{B_i}}
\end{align*}

Where:

\begin{itemize}
    \item \( \sigma_{\text{stat},B_i} \): Statistical uncertainty from the \( i^\text{th} \) background source.
    
    \item $N_{B_i}$: Number of missing $\gamma$ counts from the background source \( i^\text{th} \).
\end{itemize}

\begin{table}[h!]
\centering
\caption{Estimated statistical uncertainties in the missing $\gamma$ contributions.}
\begin{tabular}{|l|c|c|}
\hline
\textbf{Category} & \makecell{\textbf{Number} \\ \textbf{($\times 10^6$)}} & \makecell{\textbf{Statistical} \\ \textbf{Uncertainty (\%)}} \\
\hline
Total Decays & 7542.8 &  \\
\hline
Data & 690.19 & $3.5\times10^{-4}$ \\
\hline
Simulation & 565.27 & $3.1\times10^{-4}$ \\
\hline
Pile-up Contribution & 70.53 & $1\times10^{-4}$ \\
\hline
Background Contribution & 4.16 & $3\times10^{-5}$ \\
\hline
BRICC Process & 1.21 & $1\times10^{-5}$ \\
\hline
\textbf{Excess (Unaccounted)} & \textbf{49.04} &  $< 0.001\%$\\
\hline
\end{tabular}
\label{tab:MissingGammaSummary1}
\end{table}

\subsubsection{Calculation of Systematic Uncertainty}

Several sources of systematic uncertainty were evaluated for this experiment. The dominant contributions are outlined below. Unless otherwise stated, the systematic uncertainties were estimated using the cut-variation method described in Sec.~\ref{sec:statistical_systematic}.

\textbf{\emph{Coincidence window selection:}}

The choice of the coincidence window is one of the most critical factors influencing the overall systematic uncertainty of the analysis. A coincidence window of $0.5~\mu$s was selected to group events originating from a single decay, with all subsequent triggers within this interval treated as part of the same decay sequence. An inaccurate choice of this window can significantly affect the extracted missing- $\gamma$ fraction.

To estimate the associated systematic uncertainty, the coincidence window was varied to $0.4~\mu$s, $0.45~\mu$s, $0.5~\mu$s, $0.55~\mu$s, and $0.6~\mu$s, and the full analysis chain was repeated for each case. Owing to computational constraints, this study was carried out using a representative 15-minute subset of the data. The resulting spread in the number of missing- $\gamma$ events was quantified, and the standard deviation was taken as the systematic uncertainty. This corresponds to a standard deviation of 3217.15 events, or $0.058\%$, relative to a total of 5,517,417 decay events in the 15-minute data set.

\begin{table}[h!]
\centering
\caption{Event counts as a function of coincidence time window.}
\begin{tabular}{|c|c|}
\hline
\textbf{Time Window ($\mu$s)} & \textbf{Missing $\gamma$ Counts} \\
\hline
0.40 & 525088 \\
\hline
0.45 & 522753 \\
\hline
0.50 & 520570 \\
\hline
0.55 & 518348 \\
\hline
0.60 & 515918 \\
\hline
\end{tabular}
\label{tab:TimeWindowCounts}
\end{table}

\textbf{\emph{Energy threshold:}}

A similar procedure was followed to evaluate the systematic uncertainty associated with the software energy threshold. The analysis was repeated with threshold values of 60~keV, 80~keV, 100~keV, 120~keV, and 140~keV. The resulting variation in the missing- $\gamma$ counts at the end of the analysis pipeline was used to quantify the associated uncertainty. The standard deviation of the missing- $\gamma$ counts is 14,749.84 events, corresponding to $0.267\%$ of the total 5,517,417 decay events in the 15-minute data set.

\begin{table}[h!]
\centering
\caption{Event counts as a function of energy threshold.}
\begin{tabular}{|c|c|}
\hline
\textbf{Energy Threshold (keV)} & \textbf{Counts} \\
\hline
60 & 550392 \\
\hline
80 & 550080 \\
\hline
100 & 520570 \\
\hline
120 & 520073 \\
\hline
140 & 519757 \\
\hline
\end{tabular}
\label{tab:EnergyThresholdCounts}
\end{table}

\textbf{\emph{Calibration shifts:}}

Energy calibration for all detector channels was performed using an automated procedure that smooths the energy spectrum and identifies the first peak in the ${}^{46}\mathrm{Sc}$ spectrum. In some cases, the identified peak position deviates by 1--2 ADC bins from the true peak, which can affect the extracted missing- $\gamma$ yield.

To estimate the associated systematic uncertainty, the calibration peak for each channel was manually shifted by $\pm(1$--$2)$ ADC bins relative to the automatically determined position, and the full analysis pipeline was repeated for each variation. The resulting spread in the missing- $\gamma$ counts was used to determine the systematic uncertainty due to calibration shifts. The individual channel contributions were then combined using a weighted quadrature sum,
\begin{equation}
    \text{Total uncertainty} = \sqrt{\sum_i p_i \cdot \sigma_i^2},
\end{equation}
where $p_i$ is the relative probability of each channel contributing to the detected events and $\sigma_i$ is the corresponding channel-wise systematic uncertainty. The total systematic uncertainty from calibration shifts was found to be $0.21\%$.

\textbf{\emph{Variation of the detector resolution:}}

Detector energy resolution is an important input to the simulation and can influence the estimated contribution of missing- $\gamma$ events. The resolution was characterized by evaluating $\sigma/\mu$ at multiple energies using calibrated radioactive sources and fitting these values with a resolution model. Variations in the fitted parameters modify the simulated energy spectra for individual channels and can therefore affect their contribution to the missing- $\gamma$ signal.

\begin{figure}[h]
  \includegraphics[width=0.5\textwidth]{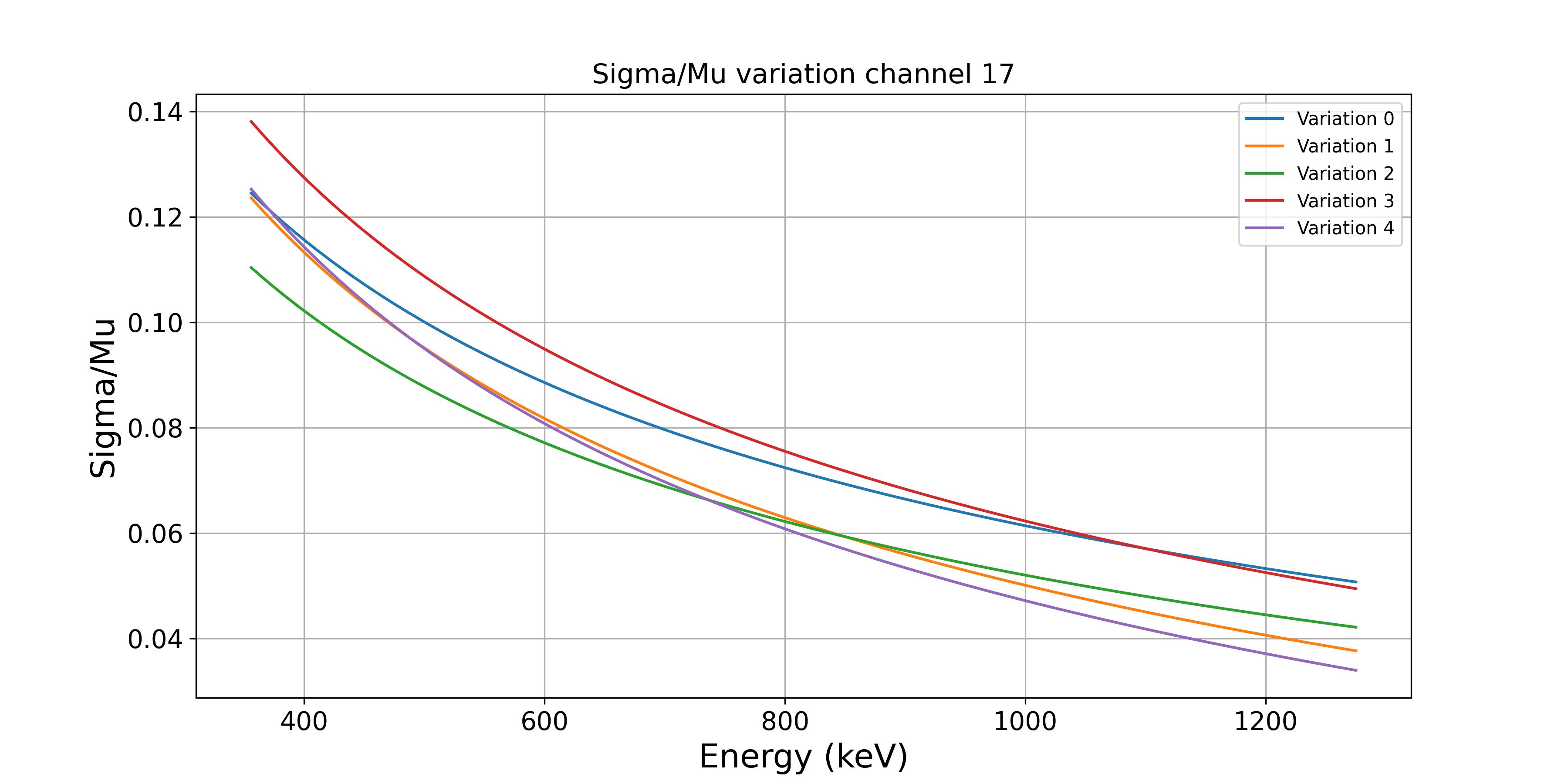}
    \caption[$\sigma/\mu$ variation and its influence on missing $\gamma$ events.]{Variation of $\sigma/\mu$ as a function of energy for a representative channel (Channel 17). Variation 0 corresponds to the nominal fit to the measured $\sigma/\mu$ data points, while the remaining variations are generated by randomly sampling the fit parameters within their covariance-defined uncertainties, yielding a set of perturbed $\sigma/\mu$ curves used to propagate the resolution uncertainty into the missing-$\gamma$ yield.}
\label{fig:Sigma_Mu_variation}
\end{figure}

To quantify this effect, the resolution-function parameters were varied by sampling Gaussian random distributions defined by the fit covariance matrix (Fig.~\ref{fig:Sigma_Mu_variation}). This procedure induces small shifts in the simulated energy response, as observed in the co-added energy spectrum for individual channels, and was repeated independently for all channels. The resulting variations in the missing- $\gamma$ counts were used to estimate the corresponding systematic uncertainty.

\begin{figure}[h]
  \includegraphics[width=0.5\textwidth]{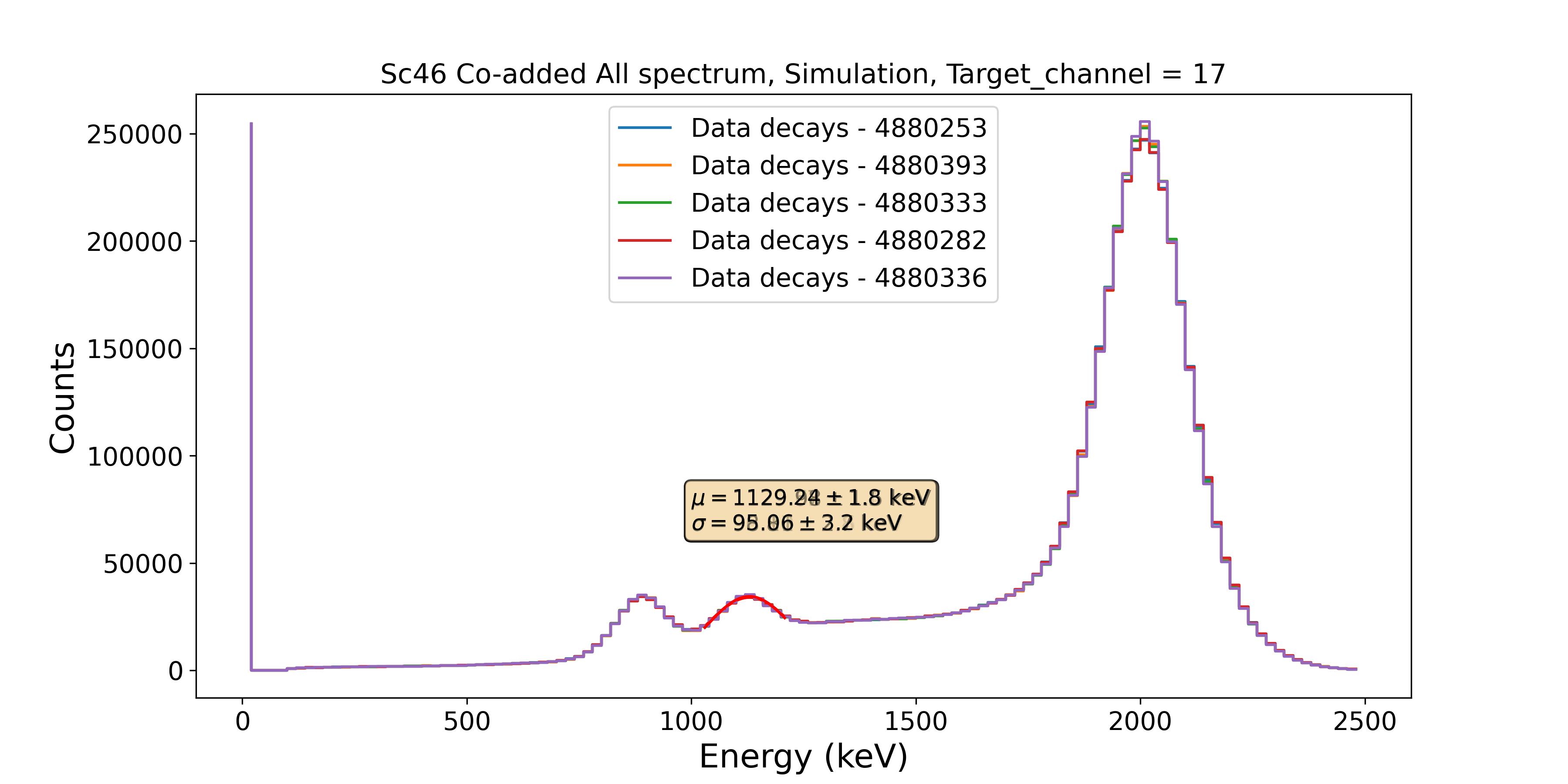}
\caption{Effect of $\sigma/\mu$ variation on the total co-added energy spectrum. Each curve corresponds to a distinct variation of the resolution function, illustrating the resulting shift in the total absorption peak and the consequent change in the reconstructed missing-$\gamma$ count. The spread across variations is used to quantify the systematic uncertainty attributed to detector energy resolution, as described in the text.}
\end{figure}

As in the previous cases, the channel-wise uncertainties were combined using a weighted quadrature sum,
\[
\text{Total uncertainty} = \sqrt{\sum_i p_i \cdot \sigma_i^2},
\]
where $p_i$ is the relative contribution of each channel to the total number of detected events and $\sigma_i$ is the systematic uncertainty in the missing- $\gamma$ yield for that channel. The total systematic uncertainty arising from detector resolution variations was found to be $0.12\%$.

The total systematic uncertainty was obtained by summing the individual contributions in quadrature, resulting in an overall systematic uncertainty of (0.373\%).

\end{document}